\documentclass[12pt,journal,onecolumn]{IEEEtran}

\usepackage{vmargin}
\setpapersize{A4}
\setmargins{2.5cm}{1.5cm}{16.5cm}{23.42cm}{10pt}{1cm}{0pt}{2cm}

\usepackage{fancyhdr}
\usepackage{color}
\newcommand{\green}[1]{\textcolor[rgb]{0,.4,0}{#1}}         % Color verd
\newcommand{\blue}[1]{\textcolor[rgb]{0,0,0.6}{#1}}         % Color blau
\usepackage{amsfonts}
\usepackage{amssymb}
\usepackage{amsbsy} % Per a \boldsymbol
\usepackage{amsmath}
\usepackage{graphicx}
\usepackage{epstopdf}
\usepackage{amsthm}
\usepackage{bbm}
\newcommand{\imag}{\mathbbm{i}}

\usepackage{array}
\newcolumntype{L}[1]{>{\raggedright\let\newline\\\arraybackslash\hspace{0pt}}m{#1}}
\newcolumntype{C}[1]{>{\centering\let\newline\\\arraybackslash\hspace{0pt}}m{#1}}
\newcolumntype{R}[1]{>{\raggedleft\let\newline\\\arraybackslash\hspace{0pt}}m{#1}}

\usepackage{pgffor} % for 
\usepackage{float} % avoid floats error latex
\usepackage{wrapfig}
\usepackage{comment}
\usepackage{algorithm}
\usepackage[noend]{algpseudocode}
\usepackage{amsmath}
\usepackage{amssymb}
\usepackage{amsthm}
\usepackage{array}
\usepackage{makecell}
\usepackage{placeins}
\usepackage{multirow}
\usepackage{url}
\usepackage{cite}
\usepackage{wrapfig,lipsum,booktabs}
\usepackage[font=small,labelfont=bf]{caption}
\usepackage{graphicx}
\usepackage{colortbl}
\usepackage[table]{xcolor}
\usepackage[all]{xy}
\usepackage{adjustbox}
\usepackage{hyperref}
\hypersetup{
    unicode=false,          % non-Latin characters in Acrobat’s bookmarks
    pdftoolbar=true,        % show Acrobat’s toolbar?
    pdfmenubar=true,        % show Acrobat’s menu?
    pdffitwindow=false,     % window fit to page when opened
    pdfstartview={FitH},    % fits the width of the page to the window
    pdftitle=tit,        % title
    pdfauthor={Camps-Valls},     % author
    pdfsubject=tit,      % subject of the document
    pdfcreator={Camps-Valls},   % creator of the document
    pdfproducer={Camps-Valls}, % producer of the document
    pdfkeywords={}, % list of keywords
    pdfnewwindow=true,      % links in new window
    colorlinks=true,        % false: boxed links; true: colored links
    linkcolor={blue},          % color of internal links (change box color with linkbordercolor)
    citecolor=blue,         % color of links to bibliography
    filecolor=blue,         % color of file links
    urlcolor=blue           % color of external links
}

\def\b{{\mathbf b}}
\def\x{{\mathbf x}}
\def\y{{\mathbf y}}

\def\k{{\mathbf k}}
\def\B{{\mathbf B}}

\def\C{{\mathbf C}}
\def\R{{\mathbf R}}
\def\X{{\mathbf X}}
\def\G{{\mathbf G}}
\def\V{{\mathbf V}}
\def\S{{\mathbf S}}
\def\Xt{\tilde{\mathbf X}}
\def\Pt{\tilde{\boldsymbol{\Phi}}}
\def\Ps{\tilde{\boldsymbol{\Psi}}}
\def\K{{\mathbf K}}

\def\Real{{\mathbbm{R}}}
\def\Integ{{\mathbbm{Z}}}
\def\Comp{{\mathbbm{C}}}
\newcommand{\HH}{\mathbbm{H}}
\title{Nonlinear PCA for Spatio-Temporal Analysis of Earth Observation Data}

\author{Diego Bueso,~\IEEEmembership{Member ~IEEE}, Maria Piles,~\IEEEmembership{Senior ~IEEE} and 

Gustau Camps-Valls,~\IEEEmembership{Fellow ~IEEE}

\thanks {---------------------------------------------------------------------------------------------------------------------------------------------------------}
\thanks{ \copyright IEEE. Personal use of this material is permitted.  Permission from IEEE must be obtained for all other users,including reprinting/republishing this material for advertising or promotional purposes, creating new collectiveworks for resale or redistribution to servers or lists, or reuse of any copyrighted components of this work in otherworks.  DOI: 10.1109/TGRS.2020.2969813}
\thanks{This work was funded by the European Research Council (ERC) under the ERC-CoG-2014 SEDAL project (grant agreement 647423), and by projects TEC2016-81900-REDT and RTI2018-096765-A-100 (MCIU/AEI/FEDER, UE). M. Piles is supported by a Ram\'on y Cajal contract (MICINN).}
\thanks{Authors are with the Image Processing Laboratory, Universitat de Val\'{e}ncia, Spain.  \url{http://isp.uv.es/}, e-mail: \url{diego.bueso@uv.es}.}
}

\begin{document}

\maketitle

\begin{abstract}
Remote sensing observations, products and simulations are fundamental sources of information to monitor our planet and its climate variability. 
%Different natural processes remain under the mixture of the  spatio-temporal signal. 
Uncovering the main modes of spatial and temporal variability in Earth data is essential to analyze and understand the underlying physical dynamics and processes driving the Earth System. 
Dimensionality reduction methods can work with spatio-temporal datasets and decompose the information efficiently. Principal Component Analysis (PCA), also known as Empirical Orthogonal Functions (EOF) in geophysics, has been traditionally used to analyze climatic data. % with different approaches as complex EOF and rotated EOF to make more interpretable decomposition and Kernel PCA to deal with nonlinear data. 
However, when nonlinear feature relations are present, PCA/EOF fails. %%%%
In this work, we propose a nonlinear PCA method to deal with spatio-temporal Earth System data. The proposed method, called Rotated Complex Kernel PCA (ROCK-PCA for short), works in reproducing kernel Hilbert spaces to account for nonlinear processes, operates in the complex kernel domain to account for both space and time features, and adds an extra rotation for improved flexibility. The result is an explicitly resolved spatio-temporal decomposition of the Earth data cube. The method is unsupervised and computationally very efficient. %%%%
We illustrate its ability to uncover spatio-temporal patterns using synthetic experiments and real data. Results of the decomposition of three essential climate variables are shown: satellite-based global Gross Primary Productivity (GPP) and Soil Moisture (SM), and reanalysis Sea Surface Temperature (SST) data. The ROCK-PCA method allows identifying their annual and seasonal oscillations, as well as their non-seasonal trends and spatial variability patterns. The main modes of variability of GPP and SM match expected distributions of land-cover and eco-hydrological zones, respectively; the inter-annual component of SM is shown to be highly correlated with El Ni{\~n}o Southern Oscillation (ENSO) phenomenon; and the SST annual oscillation is perfectly uncoupled in magnitude and phase from the global warming trend and ENSO anomalies, as well as from their mutual interactions. We provide a working source code of the presented method for the interested reader in \href{https://github.com/DiegoBueso/ROCK-PCA}{https://github.com/DiegoBueso/ROCK-PCA}. %\green{the magnitude and phase components of SST uncouple perfectly the annual oscillation from the global warming trend and the ENSO anomalies, while their mutual interactions inside the amplitude and phase spatialization.}

%\red{This paper is accompanied by the open-source package {\tt ROCKPCA}, which provides a flexible, fast, and ready-to-use implementation of the method. website? demo?}
\end{abstract}


\section{Introduction} \label{sec:intro}

%%%%%%%%%%%%%%%%%%%%%%%%%%%%%%%%%%%%%%%%%%%%%%%
% Spatio-temporal data deluge: data is not information!
%%%%%%%%%%%%%%%%%%%%%%%%%%%%%%%%%%%%%%%%%%%%%%%

In the last few decades, we have witnessed an ever growing availability of Earth system data: along with improved remote sensing observational data, a plethora of products, climate simulations and reanalysis data are now widely available. Yet, {\em data} does not necessarily mean {\em information}, and thus extracting the most important components (features) of the data is an urgent need, as well as a matter of very active research. Earth observation data is often described in both space and time (i.e. data cubes), and extracting their spatio-temporal components and patterns is one of the main goals in the geoscience and climate science communities. Such patterns are essential to {\em analyze} and {\em understand} the underlying physical dynamics and processes driving the Earth system~\cite{Guttman89,Walter2010}.

%%%%%%%%%%%%%%%%%%%%%%%%%%%%%%%%%%%%%%%%%%%%%%%
% We need to extract components automatically!
%%%%%%%%%%%%%%%%%%%%%%%%%%%%%%%%%%%%%%%%%%%%%%%

Natural processes, however, are usually masked by complex spatio-temporal feature relations, which makes the problem of identifying modes of variability specially challenging. 
%\red{An extensive high resolution spatio-temporal data for different natural variables such as surface temperature, chlorophyll density or soil moisture are available for the last 10-30 years~\cite{aaaa,bbbb,cccc}.} 
The challenge is to derive spatial and temporal components that summarize the information content of the data cubes, while being physically meaningful and interpretable. Traditional techniques of feature extraction, such as the removal of mean seasonality, temporal trends, parametric fitting or harmonic decomposition are practical and commonly used~\cite{Dong2006,Gehne2014,Brands2017,Mandic2013,Zhao2017}. However, they require prior knowledge and assumptions, and therefore impose expected relations that could not necessarily be found in data.
%\footnote{Whether of not one has to remove trends of the time-series is a debatable question. In the one hand, keeping trends can lead into mistaking spurious correlations for causal relationships since simple trends such as a linear trend can be observed for many reasons hardly link to each others. On the other hand, trends can actually be the results of a causal mechanism that we might want to infer. Other methods such as Granger Causality with an vector autoregressive model require that the time-series are stationary are at least co-integrated. Thus in some case, differentiating the time-series can be necessary.} 
This is probably the main reason why data driven decomposition methods have been largely adopted in geosciences and climate science in the last decades~\cite{forootan2016,volkov2014,bauer2013,rotatedPCA,Chen2017,Rousi2015,Mukhin2015,Dai2000}. 

%%%%%%%%%%%%%%%%%%%%%%%%%%%%%%%%%%%%%%%%%%%%%%%
% ML to the rescue!
%%%%%%%%%%%%%%%%%%%%%%%%%%%%%%%%%%%%%%%%%%%%%%%
Machine learning in general and dimensionality reduction in particular may help in extracting spatial and temporal components automatically. Dimensionality reduction methods can generally deal with data cubes and find the main features (components) efficiently. 
The application of such methods may help in uncovering relevant spatial patterns and dynamics of the underlying physics governing the Earth system. These techniques are also very useful to summarize (compress) the data into a reduced set of informative components\footnote{Here the term `information' is used loosely and refers to sensible criteria to drive dimensionality reduction methods, such as retaining most of the variance (like in PCA), correlation (like in cross-correlation analysis, CCA) or covariance (like in  partial least squares, PLS)~\cite{CampsValls09,Arenas13}}. 
The analysis of the extracted components can shed light in the understanding of the Earth system because the intrinsic components may reveal correlations with known physical processes. Indeed, dimensionality reduction is widely used in the analysis of climate dynamics and teleconnections, and it is a key first step in observational causal discovery~\cite{Rousi2015,runge2015}.

%%%%%%%%%%%%%%%%%%%%%%%%%%%%%%%%%%%%%%%%%%%%%%%
% People do this ... 
%%%%%%%%%%%%%%%%%%%%%%%%%%%%%%%%%%%%%%%%%%%%%%%

%\green{
%Climate variability are the result of exceedingly non-linear interactions between very many degrees of freedom or modes, beeing characterized by non-linearity  and  high  dimensionality~\cite{dijkstra2013}. Consequently, find ways to reduce the dimensionality of the system to find an acurate representation that show hidden processes into the spatio-temporal mixture, besides to link these modes to the dynamics/physics of the system.
Principal Component Analysis (PCA), also known in geophysics as Empirical Orthogonal Functions (EOFs), is widely used to obtain compact representations of the signal, and has been widely exploited to obtain spatio-temporal features in climatological studies~\cite{bauer2013,volkov2014,forootan2016}. Many extensions of PCA have been presented to deal with different specificities and applications in geophysics, including the extended EOF, the Multivariate EOF~\cite{Gehne2014} or the Principal Oscillation patterns ~\cite{Wang2008} (see also a review in \cite{Hannachi2007}). Interestingly, the Singular Spectrum Analysis (SSA) introduced the possibility of extracting spatial patterns at multiple time scales~\cite{Mahecha2010}, but at the cost of introducing a delay parameter which makes the decomposition very sensitive to this parameter. A number of approaches not strictly related with EOF have also been proposed, such as the ones based on temporal domain periodicities and adaptations (e.g. ~\cite{liu2012,Cremer2018}) or time-frequency transformations ~\cite{Hirsh2018,Clainche2018}. In recent years, machine learning decomposition approaches have emerged, mainly based on Gaussian Processes~\cite{Jaako2009}, Bayesian reconstruction~\cite{Mukhin2015}, and on %on more cost-effective and scalable solutions such as 
low-rank tensor learning ~\cite{Yu2015}.
%Recently, other approaches not strictly related with EOF, focused on search into the temporal domain~\cite{Cremer2018}, assuming periodic or quasi-periodic signals such as the Functional Factor analysis~\cite{liu2012}, the Dynamical Mode Decomposition~\cite{Hirsh2018,Clainche2018}, or introduce a less computational scalable decomposition in the case of Low-rank Tensor Learning decomposition ~\cite{Yu2015}. %, but in real physic problems processes  ussually are acopled or correlated, independence will fail searching in geophysic interpretation of systems.
%In recent years, some emphasis has been payed to Bayesian decomposition approaches, such as the methods based on Gaussian processes~\cite{Jaako2009} or Bayesian reconstruction~\cite{Mukhin2015}, \red{nevertheless, the computational cost involved is some cases prohibitive}. 
Most of these methods, however, assume orthogonality, periodicity, linearity or are not computationally affordable to deal with high dimensional problems. To deal with the spatio-temporal decomposition of Earth system data cubes, it is desirable that the methods i) are able to extract features which are potentially correlated --since real physical processes are usually coupled-- , ii) are able to unveil the natural nonlinear relationships present in the data, and iii) do not assume any arbitrary  parametric response function \cite{dijkstra2013}.

%% ICA dona peu a explicar per que introduim  la kurtosis ;) gastem el 4 moment ademes de la varianza
%% yeah! crec que eixa cita se la podem guardar per a la seccio de kurtosis, i aci fem una intro general nomes

%%%%%%%%%%%%%%%%%%%%%%%%%%%%%%%%%%%%%%%%%%%%%%%
%% We do this ... nice! 
%%% Our method solves the previous problems
%%%%%%%%%%%%%%%%%%%%%%%%%%%%%%%%%%%%%%%%%%%%%%%

In this paper, we introduce a nonlinear PCA based on kernel methods~\cite{CampsValls09} that addresses the main shortcomings found in the existing literature of spatio-temporal data decomposition. %However, PCA is computationally demanding with a large number of features, and at the same time the spatio-temporal components are not disentangled in the feature space.
The proposed method, called Rotated Complex Kernel PCA (ROCK-PCA for short), has the following properties:
\begin{itemize}
    \item {\em Nonlinear.} The method works in reproducing kernel Hilbert spaces to account for nonlinear processes~\cite{Scholkopf02}. Kernel methods have excelled in many problems in remote sensing, geosciences and climate sciences, mainly in classification and parameter retrieval, yet also for feature extraction and dimensionality reduction~\cite{Arenas13,CampsValls09,CampsValls11mc}.%CampsValls09wiley
    \item {\em Flexible.} An extra rotation is included to improve the flexibility and physical interpretation of the decomposition. Typically, this has been addressed by means of the Varimax rotation~\cite{kaiser1958}. Here we propose the {\em Promax} oblique method for improved versatility~\cite{promax}. This rotation method alleviates the orthogonality constraint, and makes the principal components physically interpretable.
    \item {\em Space and time decoupling.} The method operates in the complex (kernel) domain to account for space and time features~\cite{esquivel2008,complexPCA,bauer2013,forootan2016}. The  spatial and temporal modes are treated via the Hilbert transform~\cite{HilbertTransform}, thus leading to spatial and temporally explicit eigendecompositions easy to analyze. Information about the amplitude and phase of the spatio-temporal features is extracted. %\green{ and remaining a extra-dimension to explore (amplitude and phase) instead a real valued feature.}
    \item {\em Unsupervised.} The criteria of maximum projection kurtosis is implemented as an automatic means to estimate a suitable value for the three parameters of our method i.e. the kernel hyperparameters, the specific rotation and the number of extracted components. The choice of maximum tailedness of the probability distributions was previously explored in feature extraction methods based on independent component analysis (ICA)~\cite{Boergers2013}.% \green{using a higher distribution moment instead the PCA that only the explore the variance ~\cite{separationHIGHorder}}. %Independent component analysis (ICA) seeks for independent rather than uncorrelated components~\cite{Boergers2013}; this is done by retaining higher distribution moments to find the decomposition unlike PCA that only explores variance.
    \item {\em Computational efficiency.} The method is computationally very efficient. It exploits the fact that the eigendecomposition of the covariance and the Gram matrix return the same results, and hence the computational cost can be drastically reduced from quadratic ${\mathcal O}(n^2 t)$
    to linear ${\mathcal O}(t^2 n)$ in the number of pixels (grid cells $n$) and timestamps $t$  ~\cite{sharma2007}. This is of special interest in Earth sciences, where where usually $t\ll n$. The use of the Gram matrix instead of the covariance matrix is not incidental, allowing the direct kernelization of the method to derive a fast nonlinear spatio-temporal PCA~\cite{CampsValls09}.
\end{itemize}

%The method exploits the fact that the eigendecomposition of the covariance and the Gram matrix return the same results, and hence the computational cost is drastically reduced from quadratic ${\mathcal O}(n^2 t)$ ~\cite{sharma2007} to linear in the number of pixels ${\mathcal O}(t^2 n)$. The use of the Gram matrix is not incidental, and allows further improvements, such as the direct kernelization of the method~\cite{CampsValls09}, thus allowing to derive a fast nonlinear spatio-temporal PCA. However, spatio-temporal features need to represent accurately physic-like data, usually solved with the Varimax rotation ~\cite{kaiser1958}. In this work, we propose a Promax rotation ~\cite{promax} as rotation method to alleviate the orthogonality constraint and the Kurtosis parameter as optimization method to choose the best spatio-temporal representation. With this implementations, we introduce the Rotated Complex Kernel PCA (ROCK PCA) to deal with climatic spatio-temporal data decomposition.

%%%%%%%%%%%%%%%%%%%%%%%%%%%%%%%%%%%%%%%%%%%%%%%
% The paper outline
%%%%%%%%%%%%%%%%%%%%%%%%%%%%%%%%%%%%%%%%%%%%%%%

The remainder of the paper is organized as follows. In Section~\ref{sec:PCA}, we fix notation and review the different PCA-based methods, as well as the main steps needed to develop our proposed method. 
Section~\ref{sec:Rock_pca} introduces the ROCK-PCA method and its main characteristics, illustrating how the kurtosis criterion is used to automatically find the set \blue{of} the kernel parameters, specific rotation and number of principal components with a simple simulated example. 
Section~\ref{sec:experiments} shows the results of a real application to three Essential Climate Variables (ECV): global Gross Primary Productivity (GPP) from MODIS, global Soil Moisture (SM) from SMOS, and Sea Surface Temperature (SST) data from the HadISST1 data reanalysis. Conclusions and perspectives from this work are given in Section~\ref{sec:conclusions}.

%However, PCA is computationally demanding with a large number of features, and at the same time the spatial-temporal components are not disentangled in the feature representation. In this work, we propose a fast complex PCA that alleviates both problems. The complex PCA first decomposes spatial and temporal components through the Hilbert transform, thus leading to spatial and temporally explicit eigendecompositions easy to analyze. % 

\section{Principal Component Analysis Methods for Spatio-temporal Data Analysis} \label{sec:PCA}

This section reviews the main ingredients of our method for nonlinear PCA-based analysis of spatio-temporal data. After fixing the notation of linear PCA, we will describe the advantages of working in a complex domain, and of using an extra rotation transformation that makes data not necessarily orthogonal. Seeking for non-orthogonality can be useful to better meet the particular characteristics of physical variables. Then we will introduce the nonlinear extension to PCA using kernel methods, which can further enhance flexibility. % constraints ~\cite{koscielny1982} and using a rotated approach with the nonlinear version we can make a good representation of physic-like data.

\subsection{Notation and PCA}\label{subsec:notation}

Let us define a spatio-temporal data cube  $\X\in\Real^{t\times r \times c}$, defined in a $r\times c$ spatial grid and $n=r\times c$ time series observations, $\x_i\in\Real^{t\times 1}$, $i=1,\ldots,n$. 
The cube can be reshaped into matrix form as $\Xt=[\x_1,\ldots,\x_n]\in\Real^{t\times n}$, where the tilde indicates the column-wise centering operation. 
PCA serves our purpose of analyzing the feature relations contained in the data and proceeds by diagonalizing the data covariance matrix, $\C = \frac{1}{n-1}\Xt^\top\Xt \in\Real^{n\times n}$. However, given the high number of (pixel) observations, $n$, obtaining the eigenvalues and eigenvectors involves a high computational cost. An efficient alternative is to decompose the Gram matrix: $\G =\frac{1}{t-1} \Xt\Xt^\top \in\Real^{t\times t}$, which returns exactly the same solution up to a projection on the data for the first $t$ eigenvectors. This is known as the dual solution of PCA in machine learning or the $Q$-mode in statistics~\cite{bishop}. 
%Note that now we retrieve separate spatial and temporal eigenvectors, which allows us to study the signal in those terms. 
Making the eigendecomposition of the Gram matrix, we obtain the eigenvalues, $\boldsymbol{\lambda} \in \Real^{t\times 1}$, which represent the explained variance by each principal component, and the eigenvectors $\V\in \Real^{t\times t}$, which account for the directions retaining most of the variance when sorted according to $\lambda_i$, $i=1,\ldots,t$. 
It is customary to retain a subset $c$ of the top variance eigenvectors, which leads to a truncated eigenvectors matrix ${\bf V}_c\in\Real^{t\times c}$, $c\leq t$. 
Once the top $c$ components are chosen, we can use them to project the data and obtain the spatial maps of main covariation easily by $\Xt_c =\V_{c}^\top\Xt\in \Real^{c\times n}$. % and by extension the reconstructed data $\Xt=\boldsymbol{V}\boldsymbol{Xp} \in \Real^{t\times n}$.

\subsection{Complex PCA}\label{subsec:Complex_pca}
One of the shortcomings of the standard PCA approach, even in the more computational convenient dual version, is that eigenvectors and eigenvalues do not have a clear, physically meaningful interpretation in terms of spatial and temporal coordinates in the projection space~\cite{koscielny1982}. A common alternative that allows treating space and time separately is known as {\em complex} PCA~\cite{complexPCA}. The complex PCA returns a more accurate decomposition and interpretable eigenvectors for geophysical data analysis than the plain PCA version since it allows expressing the spatial and temporal components in terms of magnitude and phase~\cite{esquivel2008}.

Formally, the complex PCA applies the Hilbert transform ${\mathsf H}$ to a signal $x(t)$:
$$
{x_h(t) :=} {\mathsf H}(x(t)) = 
\frac{1}{\pi} \int_{-\infty}^{+\infty} \dfrac{x(\tau)}{t-\tau} \,d\tau.
$$
The Hilbert-transformed point is now expressed as $\x_{\mathsf H}(t) = \x(t)+\imag {\mathsf H}(\x(t))$, and hence the centered Hilbert-transformed data matrix becomes:
$$\Xt_{\mathsf H} = \Xt + \imag \Xt_h,$$
Now, one can easily demonstrate that the Gram matrix of Hilbert-transformed data reduces to
$$\G_{\mathsf H} = \Xt_{\mathsf H}\Xt_{\mathsf H}^\HH = \tilde\G + \imag\tilde\G_h\in\Comp^{t\times t},$$
{where the tilde symbol represents the column-wise matrix centering, and} we define the Hermitian of $\X_{\mathsf H}\in\Comp^{t \times n}$ as $\X_{\mathsf H}^\HH\in\Comp^{n \times t}$, and orthogonality holds, $\x(t) \perp \x_{h}(t)$, which we will use for the sake of a convenient spatial-temporal fast eigendecomposition. % and for the kernel extension. At the end, we will obtain a complex spatio-temporal features with more information than the real valued as spatio-temporal phase and modulation.

\subsection{Rotated PCA}\label{subsec:Rotated_pca}

In the rotated PCA/EOF (RPCA/REOF)~\cite{rotatedPCA} an extra rotation transformation is added. Rotated PCA is based on the Varimax rotation~\cite{kaiser1958} to maximize the Varimax criterion related with the fourth moment of probability distribution, and is implemented as a linear rotation $\B =\R\V$, where $\R$ is the rotation matrix being $\B =[\b_1,...,\b_t]$ and $\R \in \Real^{t \times t}$. An extension of the Varimax rotation is the so-called Promax rotation~\cite{promax} which introduces the transformation 
$$\b'_p=\b^p/\|\b^p\|,$$ and is applied to each component and where the power $p\in\Integ^+$ drives the components towards a ``sparse'' solution. Therefore, unlike in PCA or factor analysis, the basis now contains many zeros. %This introduces an extra hyperparameter $p$ that seeks for a more accurate representation of correlated variables. 
The Promax rotation is equal to the Varimax rotation for $p=1$. We define the Varimax rotated matrix as $\B'_p=[\b'_1,\ldots,\b'_t]$.

%\red{Two remarks follow. First, adding extra rotations after PCA has led to an immense amount of alternatives and designs\footnote{Essentially two families are found: either {\em orthogonal methods} (like equamax,  orthomax,  quartimax, and varimax), or {\em oblique rotation methods} (like binormamin, biquartimin, covarimin, direct oblimin, indirect oblimin, maxplane, oblinorm, oblimax, obliquimax, optres, orthoblique, orthotran, promax, quartimin, and tandem).}, yet in our method we will focus on the Varimax  based method for simplicity and good results. Second, strictly speaking a PCA followed by an extra rotation does no longer return `principal components', yet we will abuse language.}

%Applying it into the nonlinear complex PC's, we already obtain a final decomposition in function of kernel parameters, the number of rotated PC's and the Power of the promax rotation. When the power of promax become high, PC's tends to converge to another PC's subset where they will probably be correlated. We show in section ~\ref{sec:optimization} how to choose a good subset of this parameters. 

\begin{comment} 
\begin{equation}
\label{ec:varimax}   
Varimax=\sum_{i}^{n}(\frac{1}{t}\sum_{j}^{t}b_{ij}^4 - \frac{1}{t^2}(\sum_{j}^{t}b_{ij}^2)^2 ),
\end{equation}
 
\end{comment}

\subsection{Kernelized PCA}\label{subsec:Kernel_pca}
Working with complex and rotated PCA is often beneficial but the derived decompositions can only cope with linear feature relations. Working with nonlinear versions of PCA allows dealing with more complex data structures, avoiding the adoption of otherwise arbitrary orthogonality constraints~\cite{koscielny1982}, which is generally the case when using the rotated PCA~\cite{rotatedPCA}. %While nonlinear multivariate PCA extensions can been obtained by using neural networks, here we have the opportunity to 
Working with Gram matrices instead of covariances allows us to directly derive a nonlinear version of PCA by means of the kernel trick to derive the kernel PCA(KPCA) \cite{muller1998,Scholkopf02}.

%Alternatively, we derive here a kernelized version of the complex PCA. 
Let us define a feature map into a Hilbert space, $\boldsymbol{\phi}: \x_i\mapsto \boldsymbol{\phi}(\x_i)\in{\mathcal H}$, which is endorsed with a dot, scalar product called {\em kernel function}, $k(\x_i,\x_j) = \langle\boldsymbol{\phi}(\x_i),\boldsymbol{\phi}(\x_j)\rangle\in\Real$. The (centered) kernel matrix groups all dot products into a matrix defined as $\tilde\K = \Pt\Pt^\top \in\Real^{t\times t}$, {where the tilde represents the feature centering in Hilbert space\footnote{{Centering in feature space can be done implicitly via the simple kernel matrix operation $\tilde\K\leftarrow{\bf H}{\bf K}{\bf H}$, where $H_{ij} = \delta_{ij} - \frac{1}{n}$, $\delta$ represents the Kronecker delta $\delta_{i,j}=1$ if $i=j$ and zero otherwise.}}}. The kernel function essentially computes similarities between feature vectors. Similarly, a kernel feature vector contains all similarities between a test point $\x_*$ and all the points in the training dataset, and is defined as $\k_*:=[k(\x_*,\x_1),\ldots,k(\x_*,\x_n)]^\top\in\Real^{n\times 1}$. Then, it simply follows from the application of the Hilbert transform and the orthogonality property that the corresponding kernelized complex PCA reduces to eigendecompose 
$$\tilde\K_{\mathsf H} = \Pt_{\mathsf H}\Pt_{\mathsf H}^\HH = \tilde\K + \imag \tilde\K_h \in\Comp^{t\times t},$$
and thus we can analyze the signal in nonlinear terms, and separately in space and time components as a classic kernel PCA~\cite{muller1998}.

A bottleneck in kernel methods is the selection of the kernel function $k$. The most standard functions are polynomial and radial basis function (RBF) kernels because of their generality, ease of use and just one hyperparameter involved. In the case of working with complex algebra, however, one has to design kernel functions to deal with magnitude and phase and account for {\em circularity}. In our work we use the following complex kernel function introduced in~\cite{bouboulis2011} and further studied in~\cite{Rojo17dspkm} defined as
\begin{equation}\label{ec:complexkernel}
k_{\mathsf H}(\x,\y)= \exp\bigg(-\frac{\|\x-\y^*\|^2}{2\sigma^2}\bigg),
\end{equation}
where $\x,\y \in\Comp^{t\times 1},$ and the $^*$ is the conjugate operator. This kernel function allows us to deal with data in the complex domain, that is without losing the complex information and distribution properties, as for example the circularity.

%% Bona nit! Ho deixe ja! :)

%The kernel function can be extended as
%$$\Real (\K)=e^{-\frac{x_{R}y_{R}^T-x_{I}y_{I}^T}{2\sigma^2}}cos(\frac{x_{R}y_{I}^T}{\sigma^2})$$
%$$\mathbb{I}(\K)=e^{-\frac{x_{R}y_{R}^T-x_{I}y_{I}^T}{2\sigma^2}}sin(\frac{x_{R}y_{I}^T}{\sigma^2})$$
%where using the circular data property $x_{R}y_{I}^T=-y_{I}x_{R}^T$ we show that it is semi-positive %definite. Analyzing the each component of the complex function we see that, the real kernel function is %symmetric and have $ diagonal(\Real (\K))\geq 0$, the imaginary kernel function is skew-symmetric and have a %null diagonal $(x_{R}x_{I}^T=0)$, making semi-positive definite for circular data and hold the complex %properties of the data and of course, is hermitian $\K^{H}=\K$.\newline

\section{ROCK-PCA: Rotated Complex Kernel PCA}\label{sec:Rock_pca}

Our proposed ROCK-PCA is essentially the combination of the previous PCA-based methods. Algorithm~\ref{euclid} gives the pseudocode of the algorithm. ROCK-PCA performs the eigendecomposition of the kernel matrix in \eqref{ec:complexkernel} using data in the complex domain mapped using the Hilbert transform~\cite{HilbertTransform}, and further rotated with a Promax transform~\cite{promax}. Complex-valued processes return us more useful components, as for example, the interpretation of phase-modulation decomposition against only the real part returned by regular PCA. Kernel PCA introduces the possibility of searching for a nonlinear decomposition by mapping the original data into the Hilbert space. The Promax rotation redistributes the variance onto a more interpretable subset of components while avoiding the orthogonality constraint. Note that all parameters (number of components $c$, kernel hyperparameter $\sigma$, extra Promax rotation parametrized with $p$) are chosen by maximizing the kurtosis $\kappa$ of the Promax-rotated components.
It is important to remark that, by using specific hyperparameters, one can retrieve standard methods in the literature, such as Varimax and the kernel PCA, Varimax, and plain PCA, which demonstrates that ROCK-PCA generalizes them. For example,  using a rotation power $p=1$ and a sufficiently large $\sigma$ parameter, ROCK-PCA reduces to the Varimax rotation. Also, using only the real part of the Hilbert transform and avoiding the rotation, ROCK-PCA translates into kernel PCA. We provide source code of the presented method and a working demo in \href{https://github.com/DiegoBueso/ROCK-PCA}{https://github.com/DiegoBueso/ROCK-PCA}.
%Projection of the new data is straightforward, as well as obtaining the spatial and temporal components. These issues are further explained in what follows.

%%%%%%%%%%%%%%%%%
%%%%%%%%%%%%%%%%%

\subsection{Optimization by maximizing kurtosis} \label{sec:optimization}

%As multioutput solution method, we found a great number of possible solutions for our method that absolutely need a processes that converge in only one set of principal components. 
%We decide to use a loss function that searches for the most separable set with the proposed method restrictions, in particular, searching for subset of PC inside the nonlinear solutions and his multiple rotation for different number of PC. 

In order to select the optimal set of parameters, we maximize the kurtosis of the projections. This is a standard criterion in the ICA literature~\cite{fourthICA,separationHIGHorder,PenaCluster,PenaOutlier}, which has given good performance in a wide range of applications: from speech separation to non-stationary phase estimation and outlier cluster identification~\cite{castella_moreau_2010,Mirko2009,PenaOutlier}. Adopting the kurtosis of the projections is useful to seek for data separability, and it is a simple descriptor to account for the density shape~\cite{kurtosisMeaning}. %\green{In our case, we know that the variance of the sample mean depends on the population variance, yet the variance of sample variance, that is the definition of varimax criterion, depends on the distribution shape which is well-captured by kurtosis ~\cite{kurtosisMeaning}}.
%This in turn allows us to find inside a maximized variance group from PCA methods a subset which reach the variance maximizing the kurtosis. 
Actually the Varimax rotation is %based on the maximization of {\em varimax parameter} defined \red{as the variance of sample variance} and is 
linearly related with the kurtosis, Varimax = $\sigma^{4}(\kappa - 1)$. In our case, we maximize the kurtosis $\kappa$ to automatically set the ROCK-PCA parameters. In particular, after computing the projections {$\V_{\mathsf H}\in\Comp^{t\times c}$} for a set of parameters $(\sigma,c,p)$ [steps 3-5], we estimate its kurtosis as $\kappa=\frac{t}{c}\sum_{i=1}^{t} ({\sum_{j=1}^{c}[\V_{\mathsf H}]_{ij}^4}) / {(\sum_{j}^{t}[\V_{\mathsf H}]_{ij}^2)^2}$ [step 6]. The maximum $\kappa$ value determines the best combination of parameters, $(\sigma^*,c^*,p^*)$ [step 7]. 
Intuitively, we are seeking for parameters leading to non-Gaussian components, and as for ICA~\cite{fourthICA}, this leads to more independent components and in turn to a more compact feature representation (i.e. a lower number of more informative components are needed to describe the signal).
%Using the PC's from ROCK PCA, the optimization includes now mapping the Hilbert space ($\sigma$), the oblique rotations ($\p$) and the number of PC's to rotate ($c$) that we define as
%$$\theta=arg_{\theta}max[Kurtosis(\V(\theta)],$$
%where $\theta=[\k,\sigma,c]$. 
With this optimization the computational complexity of the method increases linearly with the number of parameters tried. %, but still it is added to the eigendecomposition complexity which remains linear. 
Smarter greedy selection of parameters could show improved speed up and will be matter of further research. %Frequently, for climatic data, $\sigma$ tend to be close to the linear case and $c$ is closer of $99\%$ of accumulative variance.

%%%%%%%%%%%%%%%%
%%%%%%%%%%%%%%%%

\subsection{Spatial and temporal components}

%% Ja dit abans... ajo arriero! :)

%In the case of the complex circular data, needed condition to retrieve interpretable signals from the real component, its properties are conserved because using the complex RBF kernel function from~\cite{bouboulis2011}, allow us to keep this properties on the nonlinear eigendecomposition, so on, the promax rotation is equal to a linear rotation, that also keep this complex properties equal, that allow us to retrieve from the real component the interpretable data.  

%One important result is, in the case of a linear separable data, frequently a problem is to search for a subset of interpretable data in a unsupervised case, spurious components with non real meaning or mixed components appear with more explained variability and real components with a lower one. Using a nonlinear kernel solution, we are allow to find a linear solution too, but finding good distributed variance obtaining a real explained variance with the subset of interpretable PC's helped with the promax rotation. This result is naturally relevant because we can obtain a significant subset of PC's  with a real explained variance that allow us to quantify the contribution of each PC separately to the whole data.
The proposed method decomposes the data across the temporal dimension and returns $c$ temporal components in $\B'_{p}\in\Comp^{t\times c}$. This allows us to project the data along time. Nevertheless, it is customary and desirable to obtain the corresponding spatial components and the explained variance by each component too. We would need to use the kernel function and the Promax rotation to do so. However, note that this is not possible even if we forget about the extra Promax rotation and work with a plain KPCA strategy. This is because when using nonlinear kernels there is no mapping between $n$ and $d$ objects in RKHS (reproducing kernel in Hilbert spaces), unlike in the linear case where there is a primal-dual equivalent solution between $\Xt_{\mathsf H}$ and its hermitic $\Xt_{\mathsf H}^\HH$. Actually, that would be only possible for symmetric data matrices, which is meaningless. A possible workaround would be to compute (big) spatial and temporal kernel matrices, decompose them, and then compute pre-images of the projected data, which is a very complex and unstable results~\cite{BakWesSch03,KwoTsa04}. 
% Imagine we compute kernel projections for arbitrary data, and other kernel projections for its transposed version, which would lead to different amount of mapped objects into the RKHS. 
%* not possible because: 1) when using nonlinear kernels there is no mapping between n and d objects in RKHS, unlike in the linear case where there's a primal-dual equivalent solution; and 2) this would only be the case for squared symmetric X, which is meaningless. Intuitively...
%* preimages as a possible approximate solution (+citas)
We alternatively propose here to use the covariance between the computed temporal components $\B'_{p}$ and the spatial data $\Xt_{\mathsf H}$, so the extracted spatial components in ROCK-PCA are defined as $\S_{c^*} = \B'^{\HH}\Xt_{\mathsf H} \in \Comp^{c^* \times n}_{p,c^*}$, which summarize the covariation between the spatial data and the nonlinear temporal components extracted by ROCK-PCA. 
\iffalse
To compute the spatial projection performed by the complex kernel PCA, let us first define $\Pt_{\mathsf H} \in \Comp^{t \times \mathcal{H}}$ as the time series mapped into the Hilbert space ${\mathcal H}$, and $\Ps_{\mathsf H} \in \Comp^{n \times \mathcal{H}}$ as the spatial samples into ${\mathcal H}$. 
Let us define the `temporal' projection ${\bf U}_{\mathcal H}=\B'_{p}^{\HH}\Pt_{\mathsf H} \in \Comp^{c \times \mathcal{H}}$ which should equal ${\bf U}_{\mathcal H}=\S\Ps_H$, where $\S \in \Comp^{c \times n}$ are the spatial projections:
$$\S\Ps_{\mathsf H} = \B'_{p}^{\HH}\Pt_{\mathsf H}.$$
Now, right multiplying times $\Ps_{\mathsf H}^\HH$ and defining the  reproducing kernel functions $\K_{ss} = \Ps_{\mathsf H}\Ps_{\mathsf H}^\HH$ and $\K_{ts} = \Pt_{\mathsf H}\Ps_{\mathsf H}^\HH$, we can write 
$$\S\K_{ss}=\B'_{p}\K_{ts},$$ 
and thus the spatial projection equation becomes
$$ \S=\B'_{p}^{\HH}\K_{ts}\K_{ss}^{-1}.$$ 
This is however computationally very costly. Nevertheless, if instead we multiply times $\Pt^\HH$, we obtain
$$\S\K_{st}=\B'_{p}\K_{tt},$$
and then since $\K_{st}$ is not squared we need to use the pseudoinverse to compute the spatial projections as
$$\S=\B'_{p}^{\HH}\K_{tt}\K_{st}^{\dag}.$$ 
\fi
%\red{projection can easily be computed using from $\boldsymbol{Xpc}=\V^{H}\Xt_{\phi}$ using:$$ \Xt_{\phi}(\x_i,\x_j) = k(\x_i,\x_j) -\frac {1}{t} \sum_{n=1}^{t} {k(\x_i,\x_n)} $$ $$- \frac {1}{t} \sum_{l=1}^{t} {k(\x_l,\x_j)} + \frac {1}{t^2} \sum_{n=1}^{t} \sum_{l=1}^{t} {k(\x_l,\x_n)}, $$calculated using only the kernel function}.

This in turn allows us to define the explained variance by each component $i=1,\ldots,c$. % as in standard EOF methods as $\sigma_{i}^{2} = \|\boldsymbol{[\S_{c^*}]_{i}}\|^{2}/\|[\B'_{p,c^*}]_{i}\|^2$.
Note that in ROCK-PCA the obtained eigenvalues do not necessarily carry information about the {\em explained variance} as in (K)PCA because the extra rotation regroups the variance inside a subset. Using our Gram matrix approach and the complex formulation, one can obtain the equivalent equation to retrieve the variance as a function of estimated time series into the Hilbert space as 
$$\sigma_{i}^{2} = \frac{\|[\S]_{i}\|^2}{ \|[\B'_{p}]_{i}\|^2},~~~i=1,\ldots,c$$
where the variance is not equal to the eigenvalues from a KPCA decomposition as it does not preserve the Hermitian properties, and we also need a normalization because the Promax rotation breaks the orthogonal property $\V_{\sf H}^{\HH} \V_{\sf H}={\bf I}$.

\begin{algorithm}[t!]
\caption{ROCK PCA.}\label{euclid}
\begin{algorithmic}[1]
\State Apply Hilbert transform: $\Xt \in \Real^{t \times n} \to \Xt_{\mathsf H} \in \Comp^{t \times n} $
\State Build Kernel matrix: $\K_{\mathsf H} \in \Comp^{t \times t}$
\For{Each set of $(\sigma,c,p)$}
\State Eigen-decomposition $\K_{\mathsf H}$ to obtain $\V_{\mathsf H} \in \Comp^{t \times t}$
\State Promax rotation onto $c$ components: $\B'_{p} \in \Comp^{t \times c}$
\State Compute kurtosis $\kappa = \kappa(\Real[\B'_{p}])$
\EndFor{\bf end}
\State Select optimal parameters: $(\sigma^*,c^*,p^*)=\text{arg}\max_{\sigma,c,p}[\kappa]$
\State Extract temporal components: $\B'_{p} \to \B'_{p,c^*} \in \Comp^{t \times c^*}$
\State Extract spatial components: $\S_{c^*} = \B'^{\HH}_{p,c^*}\Xt_{\mathsf H} \in \Comp^{c^* \times n}$
\State Compute explained variance: $\sigma_{i}^{2} = \|\boldsymbol{[\S_{c^*}]}_{i}\|^{2}/\|[\B'_{p,c^*}]_{i}\|^2$ %, $i=1,\ldots,c$
\end{algorithmic}
\end{algorithm}

\begin{figure*}[t!]
\begin{center}
\setlength{\tabcolsep}{0.0pt}
\begin{tabular}{c}
\includegraphics[width=1.0\columnwidth]{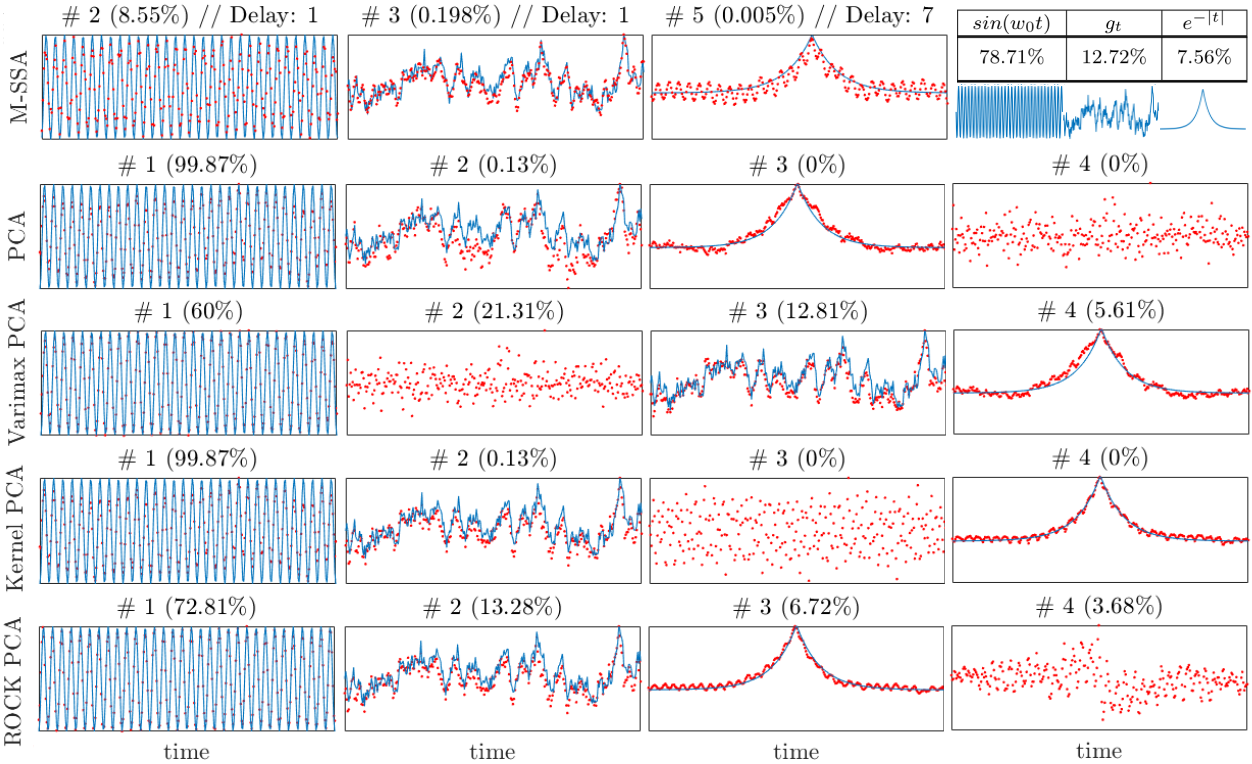} \\
\end{tabular}
\end{center}
\vspace{-0.35cm}
\caption{\label{fig:features_toy}Comparison of dimensionality reduction methods (M-SSA, PCA, Varimax PCA, kernel PCA and the proposed ROCK-PCA) in the example of section~\ref{sec:toyexample}. We show the estimated time series in red and the original signal that better correlates in blue, and indicate the explained variance (in \%, top), as well as theoretical variance per signal in the composition in the right top table. Time series without blue lines represent signals with high variance but unrelated with the original signals.}
\end{figure*}

\subsection{Synthetic spatio-temporal experiments}\label{sec:toyexample}

This section illustrates and compares ROCK-PCA to other similar methods in literature. Let us define the following toy example:
$$f(x,y,t)=e^{-|t|}cos(kr) +  g_{t}cos(ky) + sin(kxy-w_{0}t),$$
which represents a spatio-temporal signal with three additive time series ($e^{-|t|}$, $sin(w_{0})$ and $g_{t}$) and distinct spatial dynamics, and where $r=\sqrt{x^2+y^2}$, and  $g_{t}=-\alpha g_{t-1} + e_t$ is an auto-regressive (AR) model, being $k=0.5$ rad/m, $w_{0}=4.5$ rad/s and $\alpha=2$. 

\begin{figure}[H]
\begin{center}
\setlength{\tabcolsep}{0.1pt}
\begin{tabular}{cccc}
& Kernel $\sigma\times 10^4$ & Power $p$  & Components $c$   \\
\includegraphics[width=.023\columnwidth]{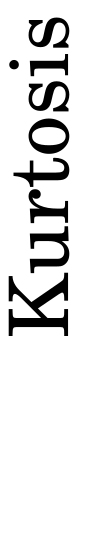} &
\includegraphics[width=.33\columnwidth]{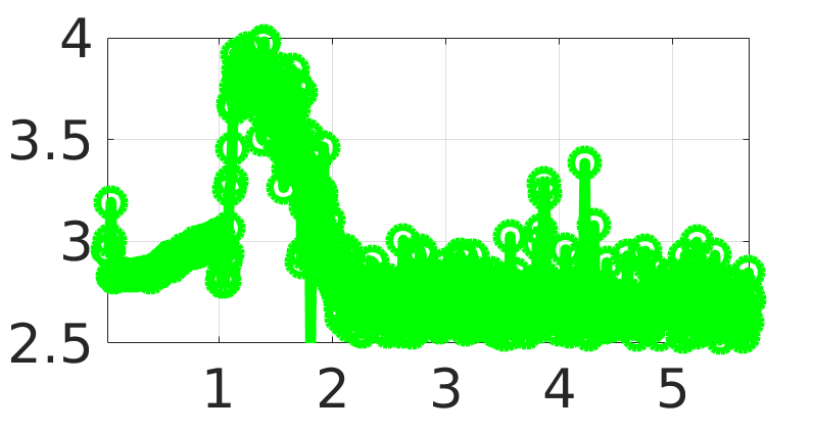} &
\includegraphics[width=.33\columnwidth]{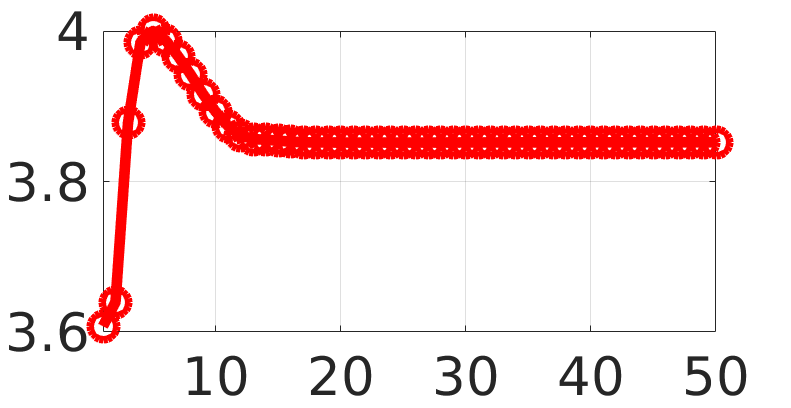} &
\includegraphics[width=.33\columnwidth]{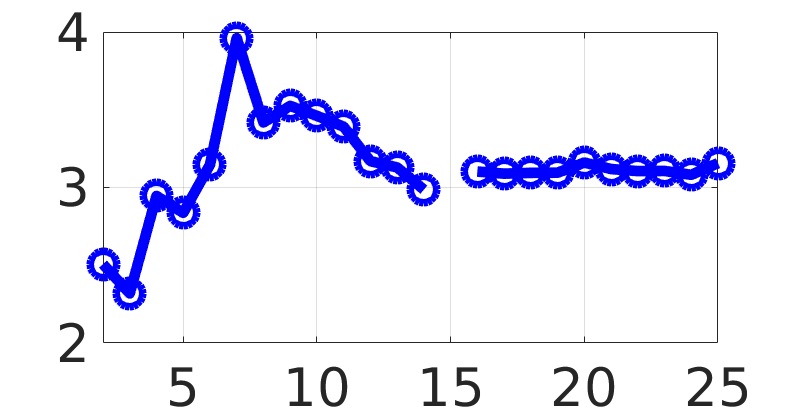} \\
%& $c$  & $k$ & $\sigma\times 10^4$ \\
\end{tabular}
\end{center}
\vspace{-0.25cm}
\caption{\label{fig:optim} Kurtosis response curve for the three method parameters: kernel hyperparameter $\sigma$, Promax power $p$ and number of extracted components $c$.}
\end{figure}

Figure \ref{fig:optim} shows the evolution of the kurtosis for the three optimization parameters in ROCK-PCA: the kernel hyperparameter $\sigma$, the Promax power $p$ and the number of extracted components $c$. Results confirm the robustness of the maximum kurtosis criterion to attain a stable solution in an unsupervised manner. In general, $c$ and $p$ response curves are less sensible than $\sigma$, and present clear local maxima. Choosing high $\sigma$ values means tending towards a linear solution as expected ~\cite{CampsValls09}. In this case, there is a maximum at a relatively small $\sigma$ value, which suggests that the problem has a nonlinear solution.

\begin{figure}[ht!]
\begin{center}
\setlength{\tabcolsep}{4pt}
\begin{tabular}{ccccc}
 & 
Original data & &
MSSA &
ROCK-PCA \\

\rotatebox{90}{\hspace{0.4cm}$sin(kxy)$} &
\includegraphics[width=0.25\columnwidth]{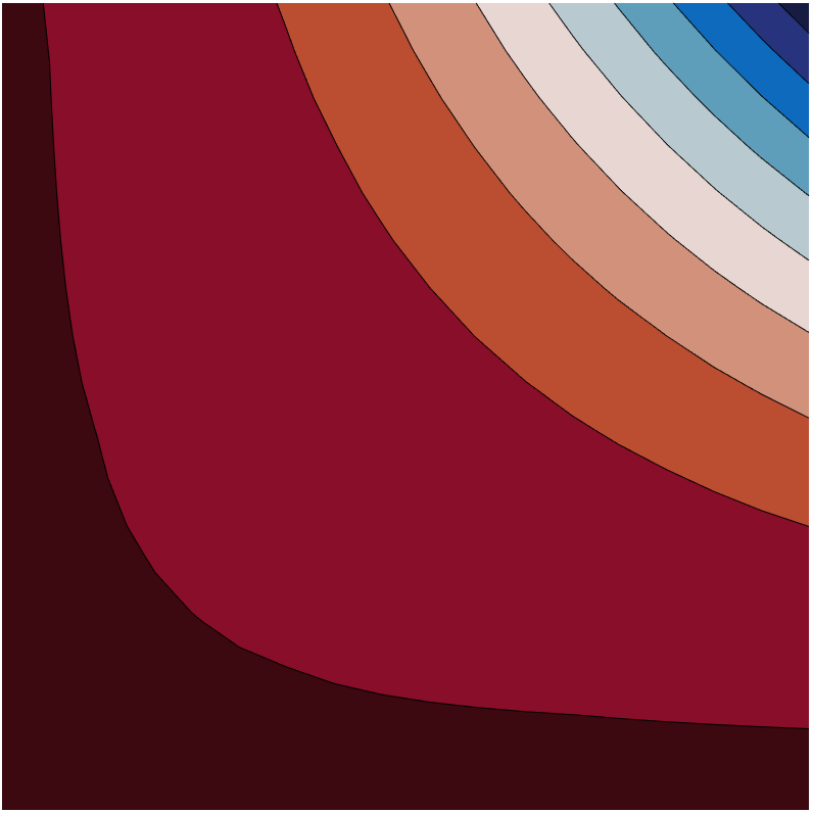} &
\rotatebox{90}{-----------------------------} &
\includegraphics[width=0.25\columnwidth]{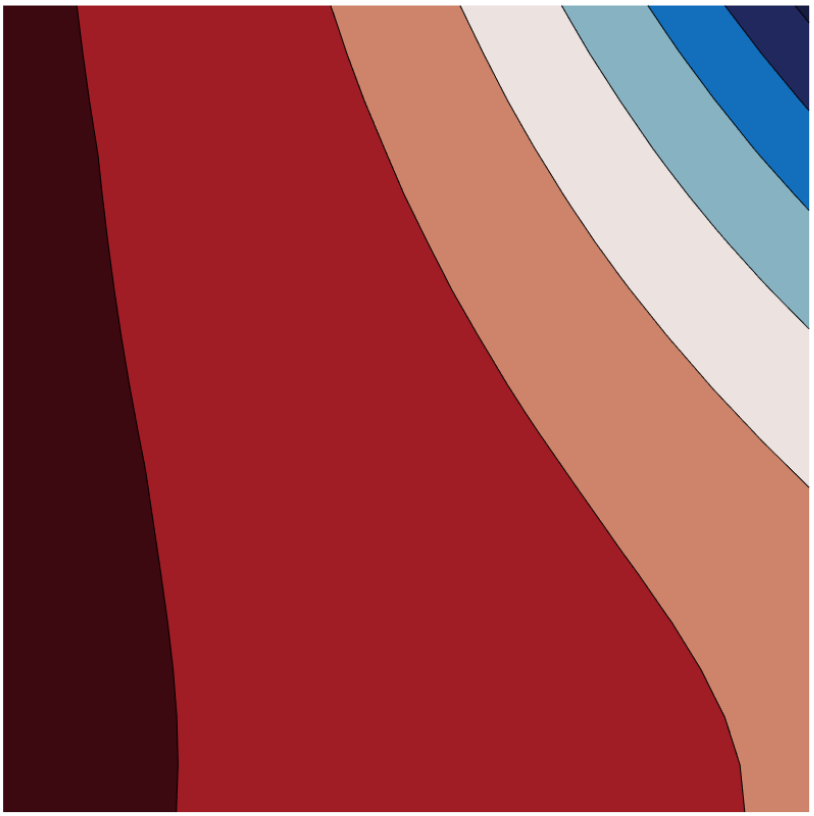} &
\includegraphics[width=0.25\columnwidth]{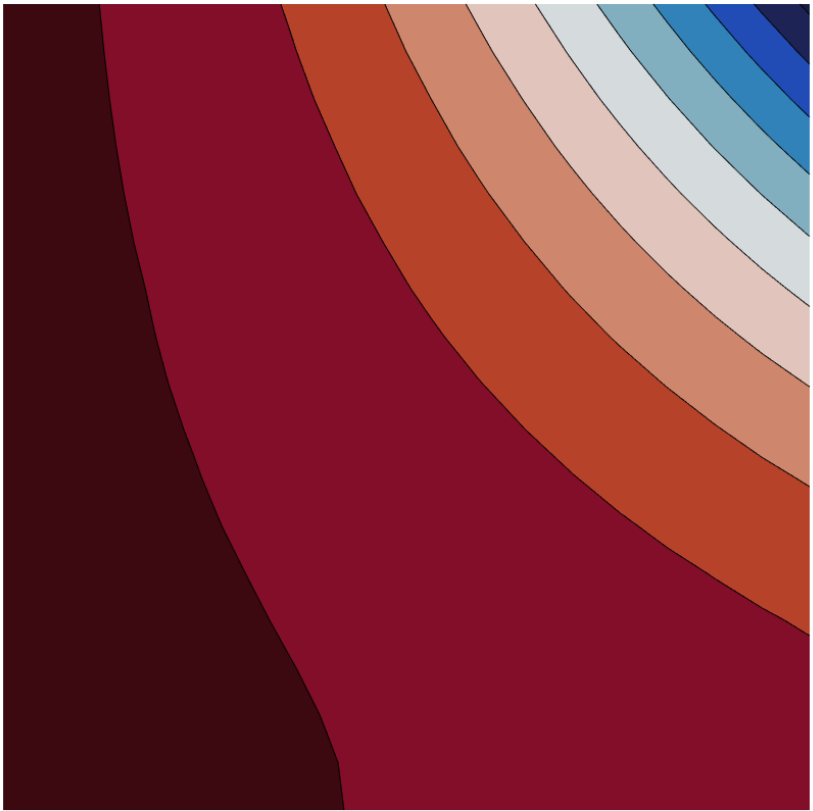} \\

\rotatebox{90}{\hspace{0.5cm}$cos(ky)$} &
\includegraphics[width=0.25\columnwidth]{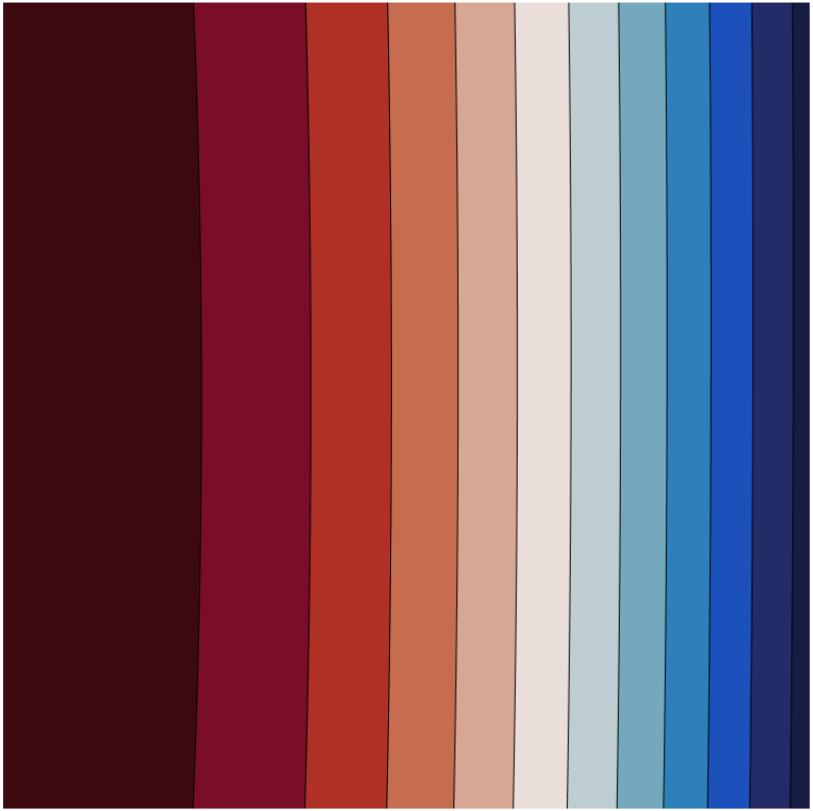} &
\rotatebox{90}{-----------------------------} &
\includegraphics[width=0.25\columnwidth]{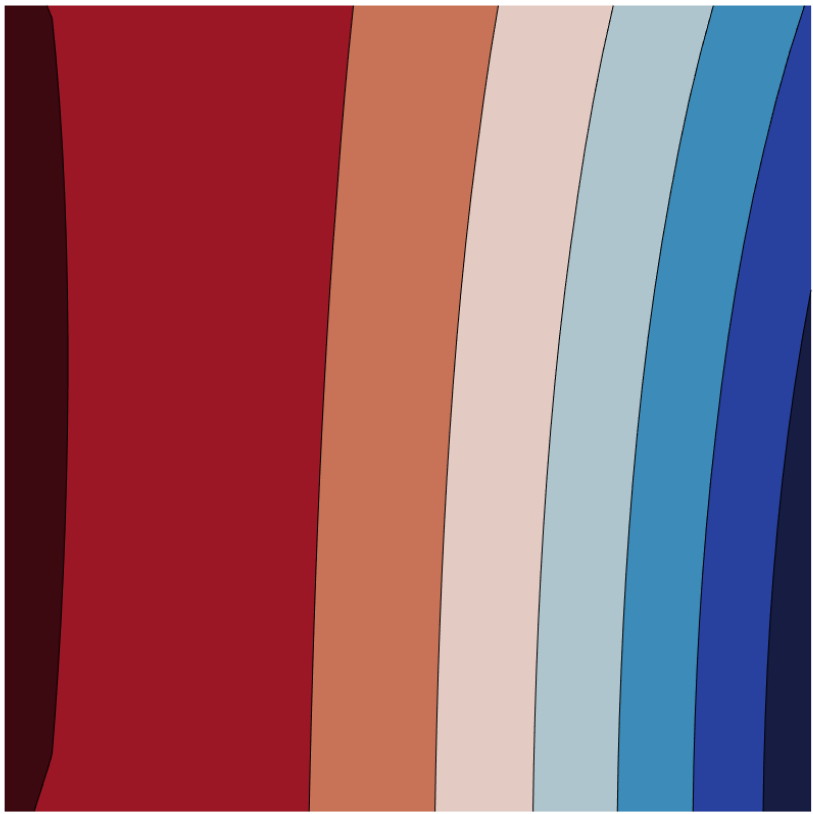} &
\includegraphics[width=0.25\columnwidth]{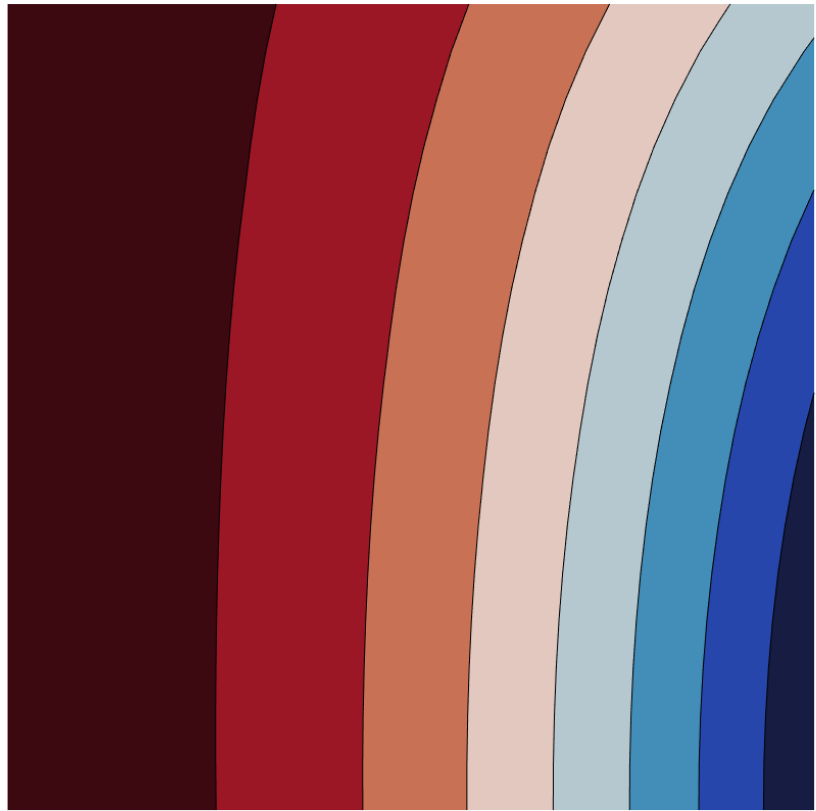} \\

\rotatebox{90}{\hspace{0.5cm}$cos(kr)$} &
\includegraphics[width=0.25\columnwidth]{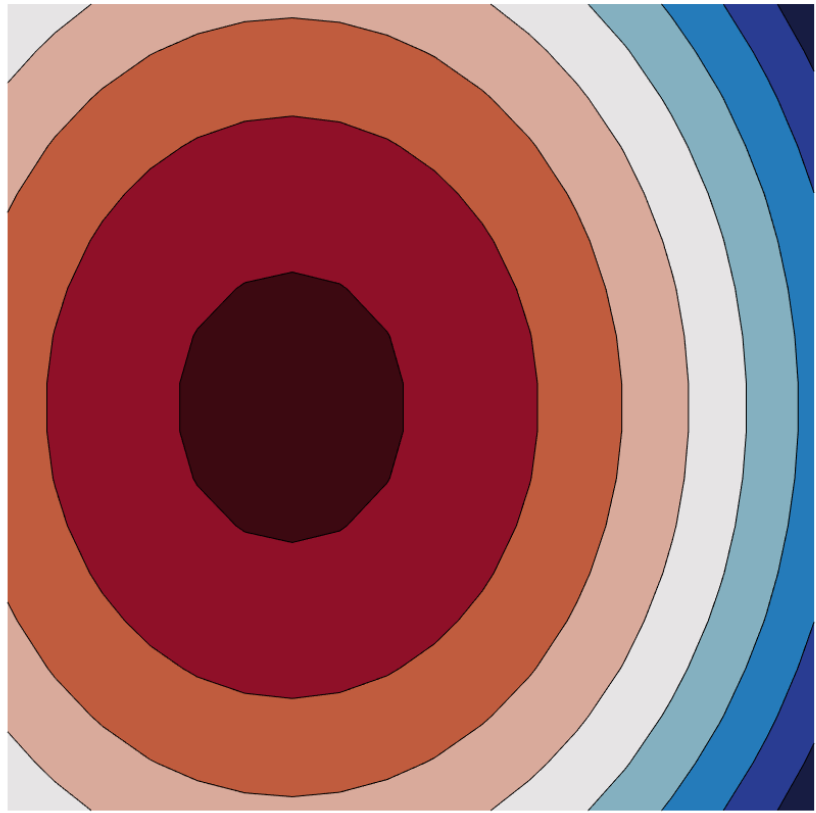} &
\rotatebox{90}{-----------------------------} &
\includegraphics[width=0.25\columnwidth]{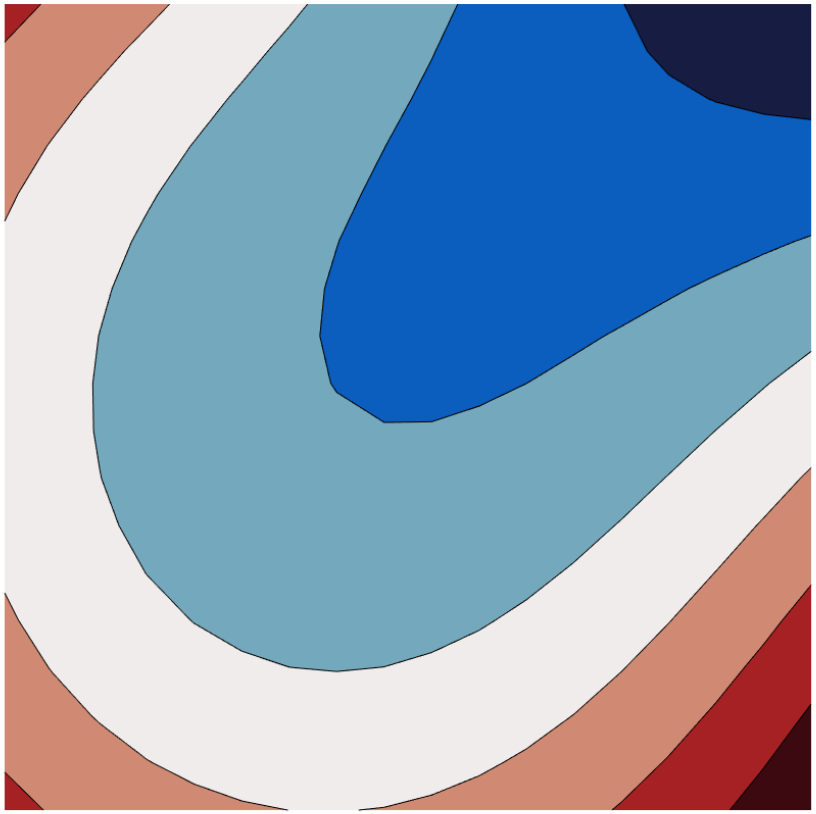} &
\includegraphics[width=0.25\columnwidth]{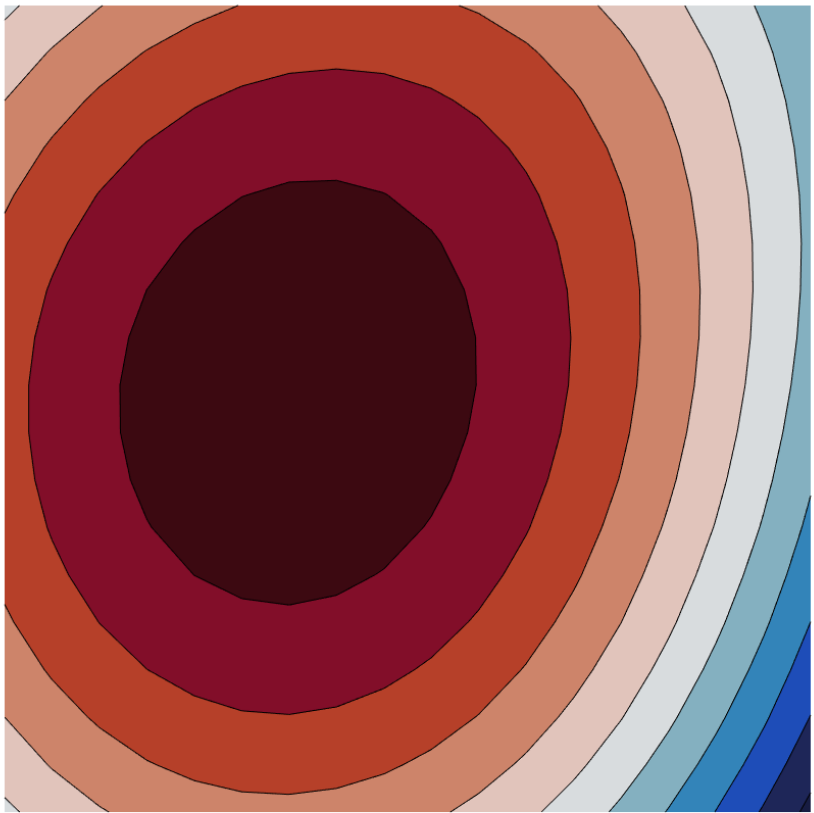} \\
\end{tabular}
\end{center}
\vspace{-0.25cm}
\caption{\label{fig:Xpc} Spatial covariation patterns from estimated time series for the original signals (left), MSSA method (middle) and the proposed method (right).}
\end{figure}

%The target of the simulated example is to extract with the proposed method each time series in separated features to show the capability of ROCK PCA to unravel mixed spatio-temporal signal to obtain his fundamental processes and redistributing correctly the variance obtaining a interpretable subset. 
Figure ~\ref{fig:features_toy} shows the results of applying PCA, varimax PCA, MSSA, kernel PCA and our proposed method ROCK-PCA. All methods are able to decompose a linear separable data set and yield components sorted according to their signal variances. Regular PCA, however, cannot find the pure sinusoidal temporal mode and therefore cannot decompose the data correctly. 
A common alternative to PCA is the addition of a Varimax rotation. Note that the solution provided by Varimax, even though it regroups variability, the second component accounts for $21.31\%$ overall variance but it is unrelated with the original signal (see table in the top right).
Multivariate Singular Spectral Analysis (MSSA)~\cite{SSA} efficiently identifies the three components, yet an intensive search over the lag parameter was needed over the second component. Results indicate that MSSA cannot reconstruct a robust set of time series with an interpretable explained variance. 
Our proposed ROCK-PCA method reproduces properly the real variances, its decomposition is not contaminated by components unrelated with the original signals (the top three components match the intrinsic components, and account for $92.81\%$ of the variance), and describes perfectly the input data. % making possible the quantification of relevance for each component for experts interpretation of real impact of each subprocesses.  

More accurate explained variance and sharper components are clearly extracted by our method, especially when nonlinear processes are involved, such as for the $sin(w_{0}t)$ and $\exp(-|t|)$ components, while no differences are observed for the simpler autoregressive process. Besides, ROCK-PCA does not need to post-process the components by lag adjustment. Figure \ref{fig:Xpc} shows the spatial covariation patterns. % obtained by data projection on the representation space. 
The ROCK-PCA patterns represent the real part of the complex estimation. ROCK-PCA spatial patterns better reflect the original spatial patterns unlike on MSSA spatial patterns.

\section{Experimental results}\label{sec:experiments}

The main goal of the proposed method is to extract information about relevant natural sub-processes that, although captured or modeled in Earth system products, are hidden under the variability of stronger modes such as the seasonal or the annual trends. The total variability of a signal \--including minor but relevant modes-- can be represented as a combination of components or modes with different temporal scales (e.g. interannual trends, seasonal resonances and anomalies) and spatial patterns. %We show here that these modes can be extracted in a smart way using nonlinear methods in the complex domain that allow treating magnitude and phase separately, and a rotation that reinforces the physical interpretation of the extracted modes. The method is unsupervised, since all the parameters are fixed with a criteria of maximum kurtosis.

In subsection A we illustrate the spatio-temporal feature extraction capability with GPP and SM data. In subsection B we show the complex spatial skills with SST data.

\subsection{Spatio-temporal variability recognition}

We here present experiments with two essential climate variables (ECVs): global Gross Primary Production (GPP) and Surface Soil Moisture (SM). We present their respective spatial and temporal decompositions with ROCK-PCA, and use the MODIS IGBP land cover classification \cite{MODISigbp} and the K{\"o}ppen-Geiger climate classification \cite{KGclass} for interpretation of results. %Results show resonances in variability decomposition and unveil spatial patterns of interest for the study of climate dynamics and its recent changes.

Both GPP and SM data sets are defined in a regular grid, and are restricted to latitudes lower than 60$^{\circ}$. The extracted components include two pieces of information: a temporal feature in the complex domain (with magnitude and phase) that accounts for the subprocesses explaining the temporal dynamic variability, and its corresponding spatial covariation which accounts for its spatial amplitude and phase. %We show the real part of the temporal with the additional phase since we are working in the complex domain. 
%, which for a  Hilbert transformation is the reconstruction of the original data from a circular complex data, and 
The spatial amplitude is useful to identify regions where the dominant mode is substantial. Using as significance threshold the median of the spatial amplitude with one positive standard deviation, we can mask the significant regions for each variability mode. The used range for the parameters is the same for all experiments. Maximum promax power $p^*$ is set on $10$ and the maximum number of components $c^*$ is set on 20. Hyperparameter $\sigma^*$ is searched in a range between 0.1 of the mean distance and 10 of maximum distance among all examples.

%Also, a short interpretation and discussion of experiment results. We compare each spatial distribution with classification land maps, extracting the relevant regions from spatial modulation fixing a threeshold in the mean of amplitude value to discuss the effectiveness of the applied method.

\subsubsection{Global GPP decomposition}\label{subsec:GPP}
Gross primary productivity (GPP) is defined as the overall rate of fixation of carbon through the process of vegetation photosynthesis. It is a key parameter for carbon cycle and climate change research. It is used to quantify the amount of biomass (Carbon mass) produced within an ecosystem over a unit of time and is usually expressed in units of $gC/m^2/day$. Quantitative estimates of the spatial and temporal distribution of GPP at regional and global scales are essential for understanding the ecosystems response to increased atmospheric CO$_2$ level, and are thus critical for sustainability and decision making. A major contribution of CO$_2$ variability comes from GPP, as the photosynthesis process is vulnerable to droughts, heatwaves, floods, frost and other types of disturbances \cite{Beer10,Jung11,Jung17nat}. Here, we consider the GPP \href{http://www.fluxcom.org/}{FLUXCOM} product, obtained by upscaling FLUXNET \cite{FLUXCOM} eddy-covariance observations by machine learning regression methods~\cite{Tramontana16bg,tramontana2015,CampsValls15GPP}. The FLUXCOM GPP product has global coverage and is provided as an eight-day composite with and spatial resolution of 5 arc-minutes ($\sim$10 Km). 

\begin{figure*}[t!]
\begin{center}
\setlength{\tabcolsep}{0.0pt}
\begin{tabular}{c}
\includegraphics[width=1.0\columnwidth]{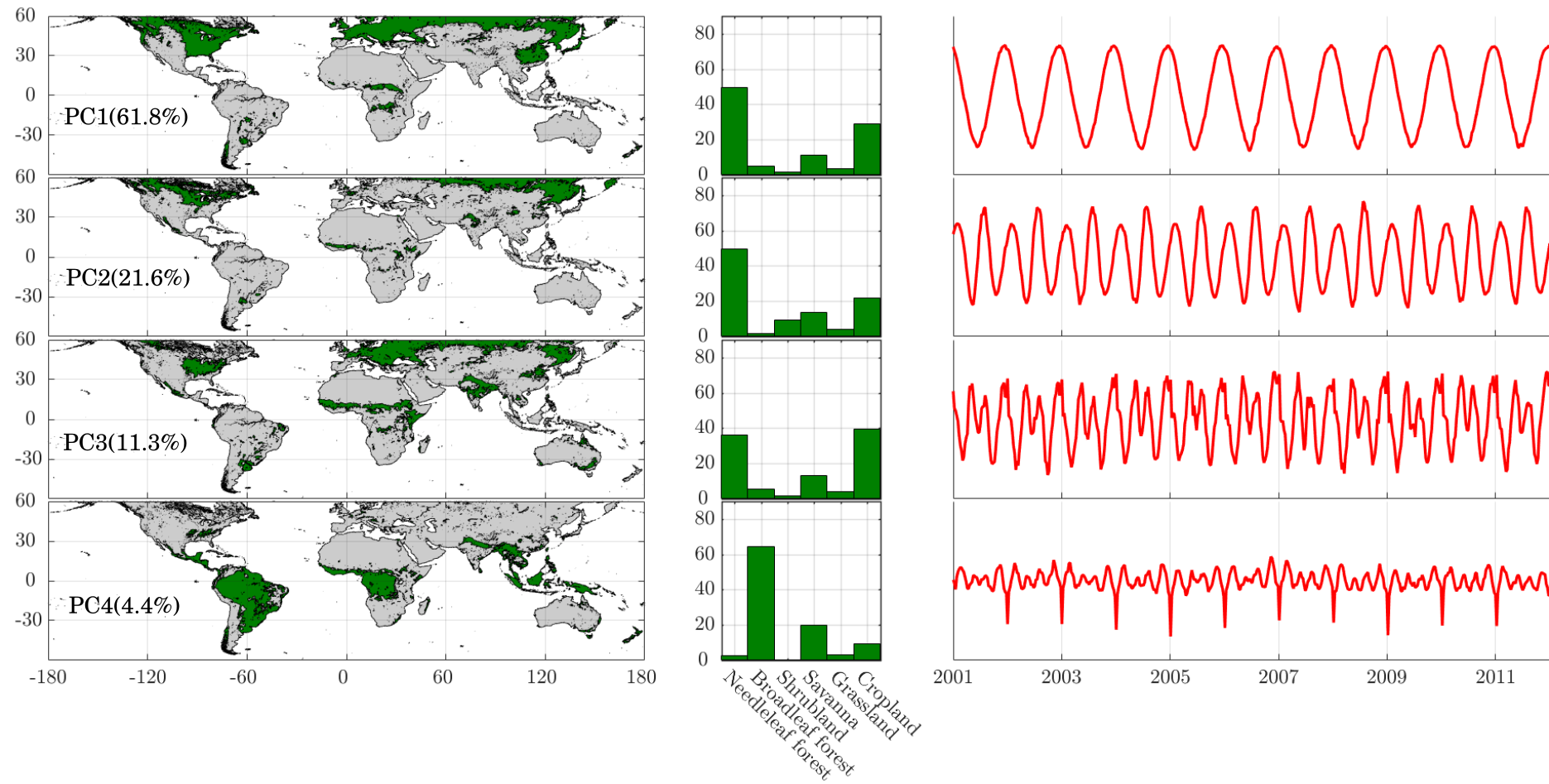} \\
\end{tabular}
\end{center}
\vspace{-0.35cm}
\caption{\label{fig:GPPfinal}Global GPP decomposition results. Left: Spatial covariation for each component, which shows the spatial distribution of the amplitude signal. Middle: Composition of each extracted components in terms of vegetation land cover classification. Right: component as representative time series of each dynamic process. Vegetation regions differentiation is shown as well, and their common dynamics as for example croplands and needle leaf forest with oscillation periods of 12, 6 and 4 months.}
\end{figure*}

Figure \ref{fig:GPPfinal} shows the ROCK-PCA decomposition results for GPP data from years 2001 to 2012. A total of six components are identified, with the 99.1\% explained variance being contained in the first four. The MODIS IGBP global land cover classification is used to spatially compare the  distribution of the extracted modes among main Earth's vegetation types. A visual examination of Figure \ref{fig:GPPfinal} reveals the annual oscillation (first component) is stronger at high latitudes 
and equatorial zones. This responds to the high variability of the freezed latitudes, which are annually coupled with the variability at the equator. %Note in the decomposition we are seeking for variance instead of mean distribution of GPP and therefore the first mode is not ~\cite{Jung11}.
%expected spatial distribution, such as annual oscillation stronger at high latitude regions and in shorter case at Equator regions, caused by high variability of freezed latitudes and and the equator, which is less variable but also annual coupled, because in the decomposition we are seeking for variance instead the mean distribution of GPP \cite{Jung11}. 
The second temporal component shows a six-month seasonality with a similar spatialization that represents subcycles of vegetation as monsoon oscillations. This result illustrates that ROCK-PCA is able to extract two different types of dynamics from similar regions, owing to the oblique rotation approach. The first and second components capture mainly the needleleaf forest variability and, to a lesser extent, the variability of certain croplands with related climatic conditions i.e. crops which phenology is characterized by a short period of Gross Primary Production stopped by the boreal/austral winter \cite{Byrne2018}. The third component covers major croplands, defined as a combination of a six-, four- and three-month periodic oscillation, resulting from the distinct growing season length of global agro-ecosystems, characterized by the crop type(s), planting and harvesting times, managerial activities and crop rotation techniques ~\cite{PEP3}. The tropical vegetation is captured by the fourth temporal component. It represents the well-known GPP variance of tropical broadleaf forests, where the peak indicates their depletion during the spring season \cite{Liang2015}. Note that inside each GPP mode, we capture all plant diversity, assuming equatorial and boreal homogeneous representation in a single dynamical mode, representing a multi-composed time series. A relevant result is that the majority of the variance (99.1\%) is represented in the top four components which cover global vegetation seasonal dynamics. Short-term and inter-annual variability is represented by the fifth and sixth components, which account for approximately the 0.9\% of the explained variance of the global GPP (not shown).

\subsubsection{Global SM decomposition}\label{subsec:SM}
\begin{figure*}[ht!]
\begin{center}
\setlength{\tabcolsep}{0.0pt}
\begin{tabular}{c}
\includegraphics[width=1.0\columnwidth]{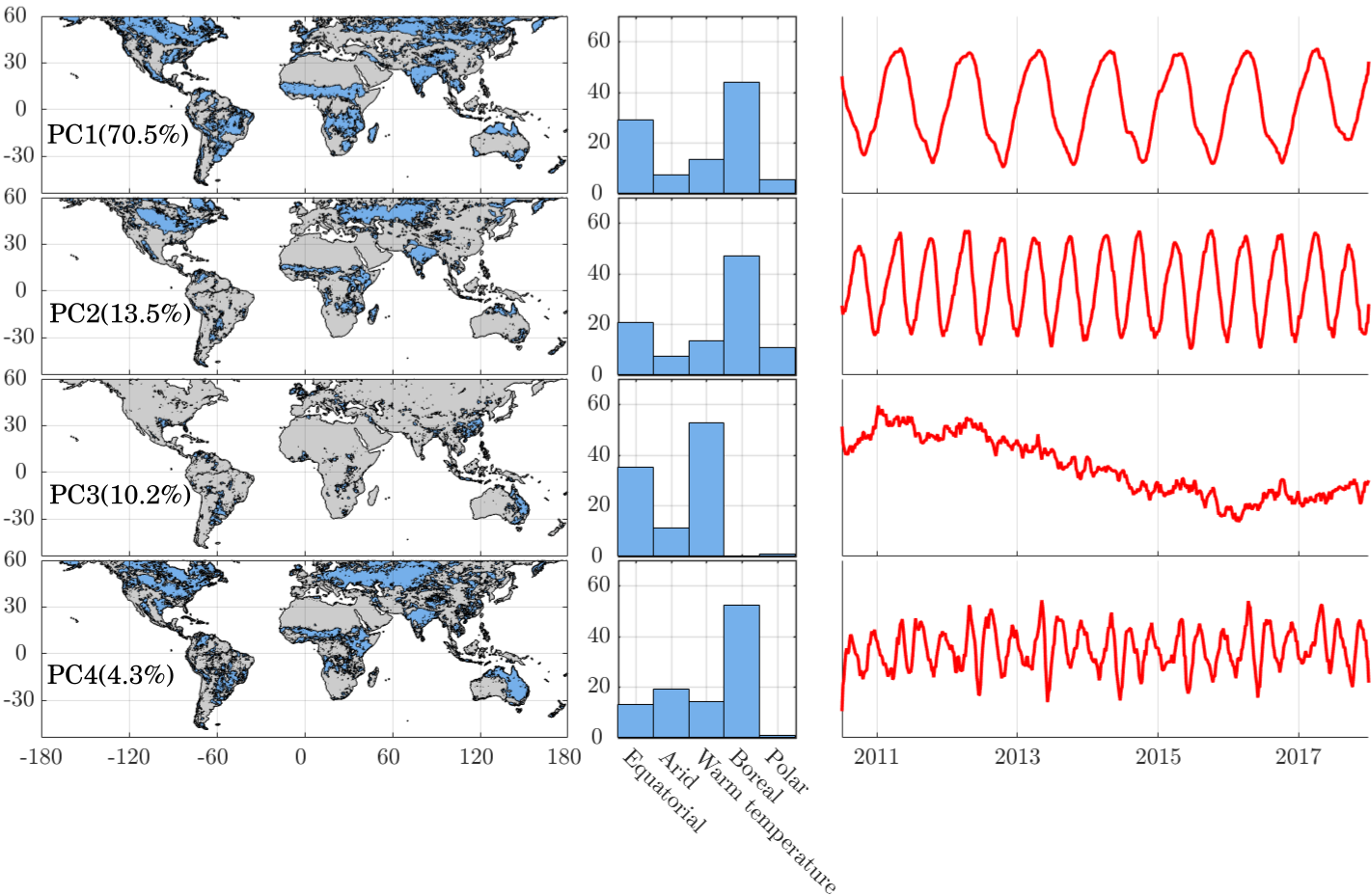} \\
\end{tabular}
\end{center}
\vspace{-0.35cm}
\caption{\label{fig:SMfinal} Global SM decomposition results. Left: Spatial covariation for each component, which shows the spatial distribution of the amplitude signal masked by significant amplitude. Middle: Composition of each extracted component in terms of KG global climate regions classification. Right: Temporal components of each dynamic process. Different time scales is show as annual, seasonal and interannual trends as well, classified in the principal climate regions.}
\end{figure*}

Figure~\ref{fig:SMfinal} shows the decomposition of the global Soil Moisture (SM) product from the \href{http://bec.icm.csic.es/}{SMOS Barcelona Expert Center (BEC)}. % and varibility decomposition in temporal modes decomposition identifing annual, seasonal and interannual relevant variance. %% GCV: ajo arriero
%Global SMOS SM data from the Barcelona Expert Center has been used in this study (http://bec.icm.csic.es/). 
Since its launch in 2010, SMOS provides global maps of the Earth's surface soil moisture (top 5 cm) every 3-days with a spatial resolution of $\sim$50 km and a target accuracy of 0.04 m$^3\cdot$m$^{-3}$. For this work, we selected the first seven years of SMOS observations, after its commissioning phase (from May 2010 to May 2017). Soil Moisture is an ECV closely related with other relevant land climate variables as surface temperature or vegetation indexes\footnote{https://www.ncdc.noaa.gov/gosic/gcos-essential-climate-variable-ecv-data-access-matrix}. Knowledge of the spatio-temporal distribution of global SM and its changes is crucial for hydrological, ecological and climate models. It links the water and energy cycles over land and is an important driver of ecosystem variability ~\cite{Seneviratne2010,Koster2011}.

The application of ROCK-PCA to the SM product yielded an optimal decomposition with seven components. The explained variance of the top four components accounts for $98.5\%$ of the overall variance, in line with a previous study focused in the spectrum of SM and its compression \cite{Katul2007}. As we show in Fig. \ref{fig:SMfinal}, the first component (70.5\%) returns an annual oscillation with a $12$-month period and a global spatial distribution, it represent the dominant mode of annual SM variability. The second (13.5\%) and fourth components (4.3\%) are cast as `resonances' of the annual SM cycle with six- and four-month periods, respectively, which correspond to subcycles of global SM as the Inter-tropical Convergence Zone (ITCZ) oscillation (second component) or the seasonal transition (fourth component). First, second and fourth components contain tropical monsoon oscillations, representing periodic sub-annual processes.

Interestingly, the third component (10.2\%) shows a non-seasonal oscillation with a large period of about $4.5$ years which can be interpreted as a long-term change in the global SM distribution. 
Comparing the spatial distribution of the amplitudes with the K\"oppel-Geiger climate classification~\cite{KGclass}, we can see that the annual and the half-year oscillation (first and second component) are distributed mostly in tropical latitudes and high latitudes, representing the oscillation of ice-covered lands. The fourth component partially contains these regions and also includes the transition between arid and wet regions. The inter-annual component is widely related with warmer (and wetter) regions representing approximately the 10.2\% of total global variability, and highlight the role of these regions as hot-spots for climate change research.  %\red{ref mahecha and Wang\cite{ref mahecha and Wang}?} 
These dry/wet regions reproduce the well-known ENSO anomalies ($2-8$ years \cite{ENSOsignal}) induced precipitation patterns~\cite{Dai2000,Lyon2005,Yeh2018}, and teleconnections \cite{Brands2017}. This remarks the close relation of satellite-based global soil moisture variability with ENSO, in line with previous research \cite{miralles2013,bauer2013,Bueso18igarss,Piles2019}. 

%and can be interpreted as the teleconnection patterns relates the impact of ENSO to global soil moisture variability 
%Likewise the GPP experiment, we can decompose the overall variability distribution in different time scales, being in the SM case the interannual variability more sensible and consequently relevant into SM global dynamics. 

\subsection{Nonlinear phase dynamics for SST and ENSO analysis}\label{subsec:SST}
%At the previous subsection we illustrate the capability of the proposed method as spatio-temporal pattern extraction way to find the different time scale dynamics as seasonal and interannual drivers and his spatial distribution presented as the amplitude of spatilized complex variable of each component. 
The proposed method works in the complex plane so magnitude as well as phase components can be extracted. We illustrate this capability with the decomposition of global Sea Surface Temperatures (SST) from the \href{https://www.metoffice.gov.uk/}{MetOffice} renalysis HadISST1 ~\cite{Parker2003}. %We show how using a complex variable we can extract more useful information than rather the real case (traditional approach) and that how using the nonlinear extraction we can show related drivers into the spatio-temporal tangled data.
We used a global SST $1^\circ$ gridded and monthly sampled cube  between years 1871-2014. 
We focus on latitudes lower than $45^\circ$ and center the analysis in tropical and middle latitudes dynamics. This is customary to avoid interference of other variables, such as the ice cover variability in high latitudes that otherwise masks the SST dynamics of middle latitudes.

\begin{figure*}[ht!]
\begin{center}
\setlength{\tabcolsep}{0.0pt}
\begin{tabular}{c}
\includegraphics[width=1.0\columnwidth]{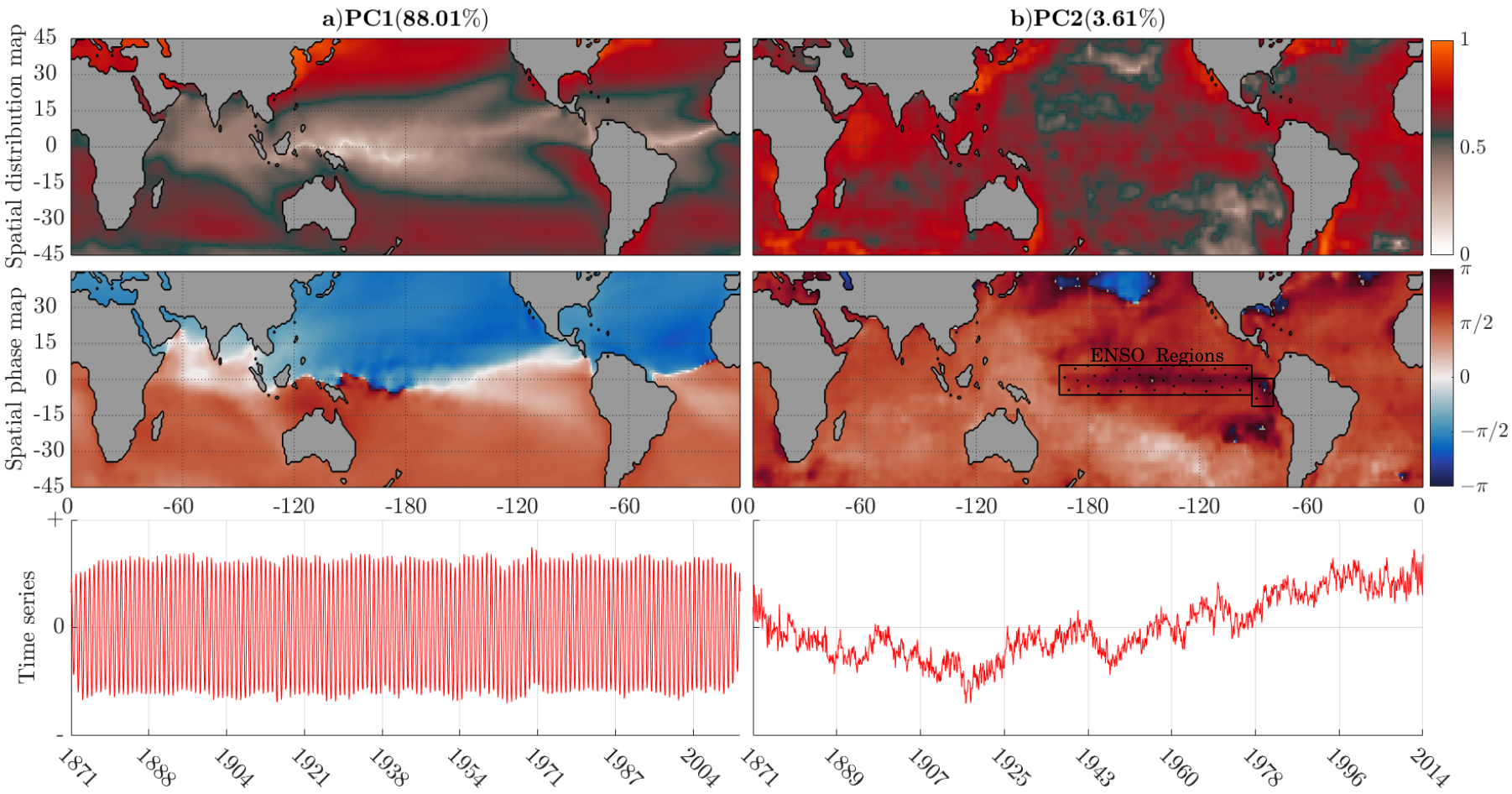} \\
\end{tabular}
\end{center}
\vspace{-0.35cm}
\caption{\label{fig:SSTfinal} Global SST decomposition results. Spatial distribution of the first (a) and second (b) components for both the spatial covariation amplitude normalized to one (top row), the spatial covariation phase (middle row), and the temporal component (bottom row). ENSO regions are also shown. }%he spatial phase of the dynamical temporal mode that represent the soft spatial differences for one dynamical mode, and on the bottom the estimated time series extracted using the proposed method. First component is related with the annual oscillation catching the 88.01\% of over all variability and the second is related with the global warming phenomenon and his phase with a almost homogeneous global distribution but with a phase that reveals where is warming, where is cooling.}
\end{figure*}

\begin{figure}[ht!]
\begin{center}
\setlength{\tabcolsep}{0.0pt}
\begin{tabular}{c}
\includegraphics[width=1\columnwidth]{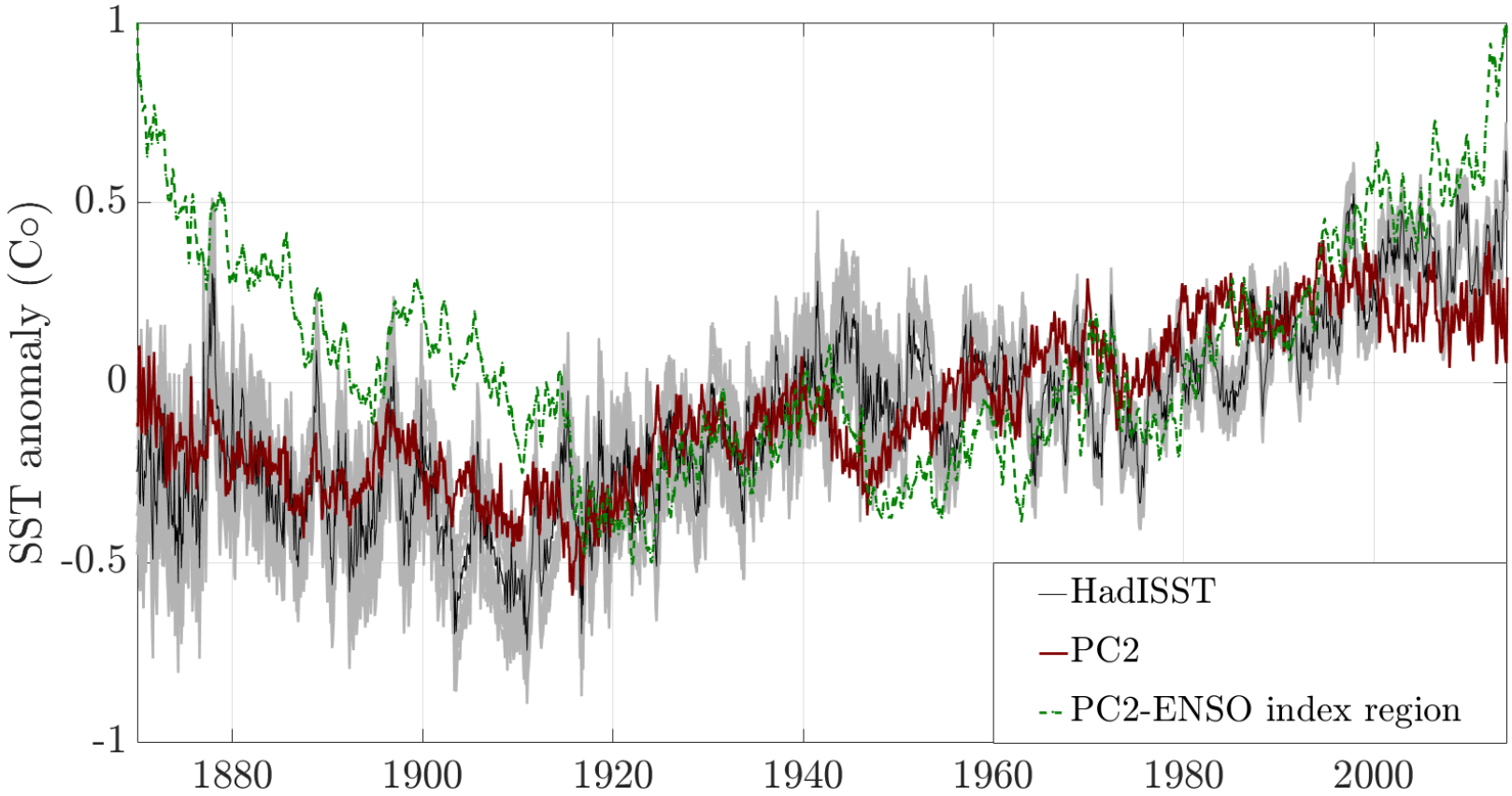} \\
\end{tabular}
\end{center}
\vspace{-0.35cm}
\caption{\label{fig:sst_trends} Comparison of inter-decadal trends of SST: mean SST Had1SST product (black line with +/- standard deviation in shaded gray), average of the second mode of variability obtained with ROCK-PCA (PC2, red line), average of PC2 extracted from ENSO index regions in Fig.\ref{fig:SSTfinal} (green line). The SST  Had1SST product is highly correlated with the total average of the second variability mode. The ENSO-region SST presents a different variability trend and a lagged response with respect to the other two.}
\end{figure}

Figure \ref{fig:SSTfinal} shows the obtained spatio-temporal decomposition in magnitude and phase. A total of five features accounted for 99.8\% of the variance. The first component shows clearly the annual north-south oscillation (88.01\%), while the second component is more related to the interdecadal variability (3.61\%). The Three following components are related to inter-annual variability as different ENSO anomalies, they are modes third to fifth representing 8.18\% variability (not shown). We focus here on the magnitude-phase characteristics of the first two components and their relation with ENSO, leaving the study of the inter-annual components for further studies. 
It can be seen that the spatial amplitude and phase of the selected components uncover extra dynamical patterns, such as for example the annual oscillation (first component)  which is represented by a two opposed-phase regions conforming a north-south oscillation boundary, representing faithfully the ITCZ line and its annual displacement (see Fig.~\ref{fig:SSTfinal}.a). The second component represents the inter-decadal temperature trend, where we can observe the recent rise of global SST with an approximate homogeneous spatial distribution that can be interpreted as the sea global warming (SGW) \cite{Grant2011}. Interestingly, the phase map shows that the ENSO region is disturbed by a positive phase in opposition to the rest of the oceans, which suggests a positive phase coupling between  ENSO and SGW. This is further analyzed in Fig.~\ref{fig:sst_trends}, representing a time dependence and, in extension, a variance dependence of ENSO events with SGW \cite{Zhang2008}. This shows well-known patterns where ENSO events variability are related with the ocean temperature raise but with a positive time delay \cite{Thomson2008,Gilbert2010,Chen2017}. Note that in negative phases (respectively, cooling regions), there is generally a low amplitude response. 

\section{Conclusions} \label{sec:conclusions}

In this paper we proposed a nonlinear dimensionality reduction method for spatial-temporal analysis of Earth observation data. The proposed method is based on kernel methods to deal with nonlinear processes and feature relations, it operates in the complex kernel domain to account for both space and time features, and adds an extra rotation that makes the components non-orthogonal to allow recovering correlated features. The method is also very efficient computationally since it can work in the dual space, which is convenient in the usual case where the amount of available pixels is larger than the number of temporal observations. If we encounter the contrary case, the formulation could be easily adapted to work in the primal for efficiency.
%\red{The proposed method allows a spatially-explicit and temporally-resolved decomposition of the main modes of variation in Earth system data. The components are nonlinear,  physically interpretable, and the extraction is fast and unsupervised.}
The method contains three parameters to tune: kernel parameter, shape of the rotation transform, and number of components to extract. To make the method unsupervised and less sensible to their selection, we proposed the optimization of the fourth moment (kurtosis) of the distribution of projections, following similar motivations in ICA approaches to signal decomposition. 

We showed performance in synthetic experiments, and three real data cubes involving land and ocean applications: global GPP, SM, and SST. The method allows identifying in a general way, annual and seasonal oscillations, as well as their non-seasonal trends and spatial variability patterns. The main modes of variation of GPP and SM are shown to match expected distributions of land-cover and eco-hydrological zones, respectively; GPP decomposition represents faithfully the principal vegetation land cover dynamics in a compressed way; the spatialization of the inter-annual component of SM reproduce accurately the global ENSO teleconnection patterns and possible novel dry/wet patterns; the SST annual oscillation is perfectly uncoupled in magnitude and phase from the global warming trend and ENSO anomalies, showing his mutual interaction as ENSO and global warming trend coupled system.

%The newly proposed feature extraction method for spatio-temporal data in remote sensing data sets is convincing in many respects. First, the method was able to ... 
One of the current limitations is that the input data set has to be sampled on a regular grid in space and time, so it cannot properly performed with gaps in the data. Some alternatives exist in the literature to resolve this, such as gap-filling the data, missing-data PCA methods, or more recent approaches based on graphs. Given our kernel-based approach, replacing the kernel matrix with a graph Laplacian would allow to resolve this problem in an elegant way. Recent literature has actually combined Laplacian eigenmaps with Takens embedding for spatio-temporal data analysis~\cite{Szekely2016}, which we will explore in the future too. Interestingly, note that ROCK-PCA extracts meaningful features even without a time embedding. 

Perhaps the most important limitation is about the  interpretability of the results, as it is definitively challenging to identify first the number of components which describe the data well and then to assign them to particular physical processes and events. While we propose here the use of kurtosis as a sensible criterion for the first, the physical interpretation (and eventual teleconnections) of the extracted components is still an unsolved problem and matter of current and active research. 

It is acknowledged that the method is general enough to work with arbitrary spatio-temporal data. The method is applicable to all kind of variables, and generalizable to work with multiple variables, not just a single one ~\cite{ multivariate_EOF}. We foresee a wide range of applications to exploit the gridded information in Earth cube initiatives, such as the Earth System Data Cube (ESDC)\footnote{\href{http://earthsystemdatacube.net/}{http://earthsystemdatacube.net/}}. 
Spatio-temporal data structures are not only encountered in Earth sciences. We anticipate applications of the method in many other fields: from epidemics, neurosciences, social sciences to economics.

\bibliographystyle{IEEEtran}
\bibliography{bibliography}

\iffalse
\input{authorsBibliography.tex}
\fi
\end{document}